%% file: EXO-13-001_temp.tex
\pdfoutput=1

\documentclass[11pt,twoside,a4paper,cmspaper,final,collab]{cms-tdr}
\def\svnVersion{418744P}\def\cmsCernNoTag{CERN-EP/2016-187}\def\cmsCernDate{\today}\def\cmsMessage{Published in Physics Letters B as \href{http://dx.doi.org/10.1016/j.physletb.2017.01.073}{\doi{10.1016/j.physletb.2017.01.073.}}}

\begin{document}\cmsNoteHeader{EXO-13-001}

\hyphenation{had-ron-i-za-tion}
\hyphenation{cal-or-i-me-ter}
\hyphenation{de-vices}
\RCS$HeadURL: svn+ssh://svn.cern.ch/reps/tdr2/papers/EXO-13-001/trunk/EXO-13-001.tex $
\RCS$Id: EXO-13-001.tex 377353 2016-12-14 17:46:43Z glandsbe $
\newlength\cmsFigWidth
\ifthenelse{\boolean{cms@external}}{\setlength\cmsFigWidth{0.85\columnwidth}}{\setlength\cmsFigWidth{0.4\textwidth}}
\ifthenelse{\boolean{cms@external}}{\providecommand{\cmsLeft}{top\xspace}}{\providecommand{\cmsLeft}{left\xspace}}
\ifthenelse{\boolean{cms@external}}{\providecommand{\cmsRight}{bottom\xspace}}{\providecommand{\cmsRight}{right\xspace}}
\ifthenelse{\boolean{cms@external}}{\providecommand{\cmsLLeft}{Top\xspace}}{\providecommand{\cmsLLeft}{Left\xspace}}
\ifthenelse{\boolean{cms@external}}{\providecommand{\cmsRRight}{Bottom\xspace}}{\providecommand{\cmsRRight}{Right\xspace}}
\providecommand{\NA}{\ensuremath{\text{---}}\xspace}
\cmsNoteHeader{EXO-13-001}

\providecommand{\PSH}{\ensuremath{\widetilde{\cmsSymbolFace{H}}}\xspace}

\title{Search for new phenomena in events with high jet multiplicity and low missing transverse momentum in proton-proton collisions at $\sqrt{s} = 8$\TeV}

\date{\today}

\abstract{
A dedicated search is presented for new phenomena in inclusive 8- and 10-jet final states with low missing transverse momentum, with and without identification of jets originating from b quarks. The analysis is based on data from proton-proton collisions corresponding to an integrated luminosity of 19.7\fbinv collected with the CMS detector at the LHC at $\sqrt{s} = 8$\TeV. The dominant multijet background expectations are obtained from low jet multiplicity control samples. Data agree well with the standard model background predictions, and limits are set in several benchmark models.
Colorons (axigluons) with masses between 0.6 and 0.75 (up to 1.15)\TeV are excluded at 95\% confidence level. Similar exclusion limits for gluinos in $R$-parity violating supersymmetric scenarios are from 0.6 up to 1.1\TeV. These results comprise the first experimental probe of the coloron and axigluon models in multijet final states.}

\hypersetup{%
pdfauthor={CMS Collaboration},%
pdftitle={Search for new physics in events with high jet multiplicity and low missing transverse momentum in proton-proton collisions at sqrt(s) = 8 TeV},%
pdfsubject={CMS},%
pdfkeywords={CMS, physics, exotics, jets, coloron, axigluon, RPV, susy}}

\maketitle

\section{Introduction}

Searching for new phenomena in final states with jets has been a tradition at hadron colliders that has continued at every new energy frontier. New particles decaying into jets are likely to be strongly produced; therefore, one expects these phenomena to have relatively large cross sections and be detectable at the CERN LHC. The challenge of searching in high-multiplicity all-hadronic final states without significant missing transverse momentum is the handling of the overwhelming multijet background. To remedy this situation, the focus of these searches has been on resonances having either narrow widths or large masses, in order to enhance the signal-to-background ratio and, consequently, the sensitivity. The presence of new phenomena in the simplest of multijet final states, the dijets, has been sought in proton-antiproton collisions at $\sqrt{s} = 0.63$\TeV by the UA1~\cite{UA1} and UA2~\cite{UA2-1,UA2-2} Collaborations at the CERN S$\Pp\Pap$S, and at $\sqrt{s} = 1.8$ and 1.96\TeV by the CDF~\cite{CDF00,CDF0,CDF1,CDF2,CDF3,CDF4} and D0~\cite{D0-1,D0-2,D0-3} Collaborations at the Fermilab Tevatron, as well as  in proton-proton collisions at $\sqrt{s} = 7$, 8, and 13\TeV by the ATLAS~\cite{ATLAS1,ATLAS2,ATLAS3,ATLAS4,ATLAS5,ATLAS6,ATLAS7,ATLAS8,ATLAS9} and CMS~\cite{CMS1,CMS2,CMS3,CMS4,CMS5,CMS6,CMS7,CMS8,CMS9,CMS10,CMS11} Collaborations at the LHC.

The complexity of these searches increases dramatically with the increase in jet multiplicity, even when several resonances are present in the production and decay chain. The reason is an exponentially increasing combinatorial background, which makes it virtually impossible to take advantage of either two-jet or multijet resonances in multijet final states. For example, consider the production of a pair of new particles X, each of which decays to a pair of particles Y that further decay into a dijet final state each. One would then expect invariant masses of four dijet combinations in the 8-jet final state to peak at the mass of the Y particle, and, likewise, the invariant masses of two 4-jet combinations would peak at the mass of the X particle. These measurements would make it seemingly easy to discern the signal from the multijet background, which lacks such features. Nevertheless, the total number of ways to arrange 8 jets into four pairs is $7!! \equiv 7 \times 5 \times 3 \times 1 = 105$, and, on top of this, there are $C^2_4/2 = 3$ possible arrangements of four pairs into two quadruplets, yielding 315 possible combinations. In practice the correct jet assignment is thus overwhelmed by the wrong combinations. Moreover, the number of ways to partition 8 jets into two quadruplets is $C^4_8/2 = 35$, so even trying to find the correct 4-jet combinations is a daunting task. Initial- and final-state radiation, as well as jet merging, makes the identification of correct pairing even more challenging.

Consequently, searches in exclusive multijet final states have thus far only been performed in the 4-jet (by ATLAS~\cite{ATLAS-4j-1,ATLAS-4j-2}, CMS~\cite{CMS-4j-1,CMS-4j-2}, and CDF~\cite{CDF-4j}) and 6-jet (by CDF~\cite{CDF-6j}, CMS~\cite{CMS-6j-1,CMS-6j-2,CMS-6j-3}, and ATLAS~\cite{ATLAS-6j}) final states. In most of these analyses no attempt was made to find the correct jet combination. Some of these analyses tag jets originating from \PQb quarks (\PQb-tagged jets), which allows them to improve sensitivity to new particles decaying to b quarks, thanks to a significant reduction in the multijet background.

In addition to these searches, a separate class of searches has been conducted in high-multiplicity inclusive jet final states. This includes searches for semiclassical black holes~\cite{dl,gt} pioneered by CMS~\cite{CMSBH1,CMSBH2,CMSBH3} and recently also conducted by ATLAS~\cite{ATLASBH1,ATLASBH2}, as well as an ATLAS search~\cite{ATLAS-multijet} for pair-produced gluinos, each decaying into either three or five jets, which appear in certain $R$-parity~\cite{R-Parity} violating (RPV) supersymmetric (SUSY) models. In these analyses, no attempt is made to reconstruct the invariant mass of jet combinations, and the analyses either use global variables, such as the total scalar sum of transverse momenta of all jets in the event, \HT~\cite{CMSBH1,CMSBH2,CMSBH3,ATLASBH1,ATLASBH2}, or the sum of reconstructed jet masses~\cite{ATLAS-multijet}; or simple counts of the total number of jets, as well as the number of \PQb-tagged jets~\cite{ATLAS-multijet}.

\section{Analysis strategy}

The first CMS black hole search in multijet final states~\cite{CMSBH1} introduced a novel technique that relies entirely on data to predict the
dominant multijet background. The technique is based on the observation that in multijet events the additional energetic jets (beyond the $2 \to 2$ hard scattering) are produced mainly via final-state radiation (FSR). This process approximately conserves the \HT in the event, while increasing the jet multiplicity. Therefore, one could use the \HT spectrum at lower multiplicities, not contaminated by potential signal, to predict its shape at higher multiplicities. This technique, subsequently used in other CMS publications~\cite{CMSBH2,CMSBH3,Stealth}, forms the basis of the present analysis.

This Letter presents the results of the first search for new physics in high-multiplicity all-hadronic final states with low missing transverse momentum that utilizes a simple kinematic analysis of the multijet final state, both in a flavor-blind analysis and in an analysis that requires at least one of the final-state jets to be b tagged. This approach makes the analysis sensitive to a large class of models of new physics. We illustrate this by considering three specific models resulting in such final states: pair production of colorons~\cite{Colorons0,Colorons1,Colorons2,Colorons3}; axigluons~\cite{Axigluons}; or gluinos decaying via RPV interactions~\cite{RPVModel}.

\begin{figure}[htb!]
 \centering
 \includegraphics[width=0.45\textwidth]{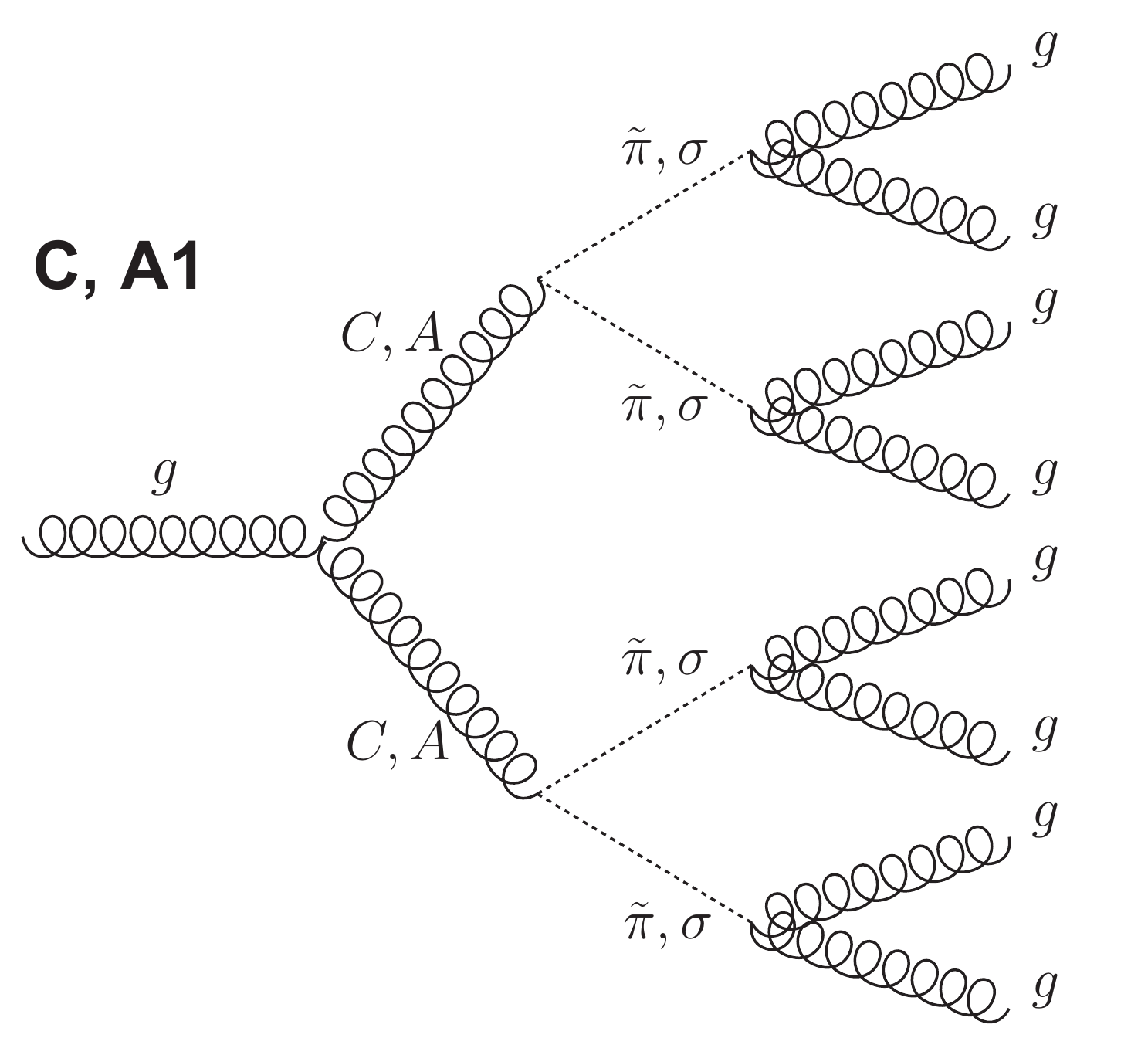}
 \includegraphics[width=0.45\textwidth]{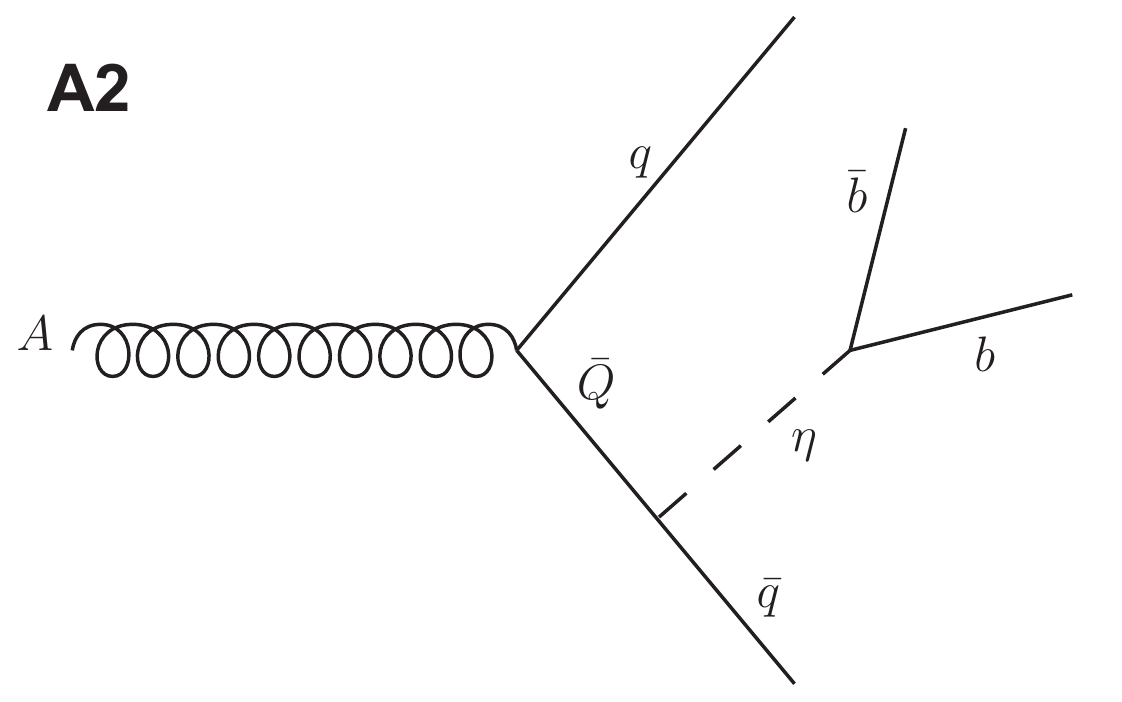}
\caption{\cmsLLeft: the dominant Feynman diagram representing the $s$-channel pair production of color-octet vector bosons, subsequently decaying into spin-0 particles and finally to gluons. The vector bosons can be colorons C or axigluons A, while the spin-0 particles can be pseudoscalar hyperpions $\tilde{\pi}$ or scalar particles $\sigma$. This process corresponds to models C and A1. \cmsRRight: the second decay mode of an axigluon considered in this analysis (A2), involving a heavy quark Q and a pseudogoldstone boson $\eta$ with Higgs-like couplings.}
\label{fig:diagram}
\end{figure}

For colorons C, which are vector color-octet particles, we consider a specific model~\cite{Dicus1,Dicus2}. In this model, strongly produced coloron pairs each decay into a pair of color-octet hyperpions $\widetilde\pi$, the lightest narrow bound states predicted in models with new strong dynamics. Each hyperpion further decays into gluon pairs, thus resulting in an 8-jet final state, as shown in Fig.~\ref{fig:diagram} (\cmsLeft). This is the most important production diagram, accessible either from a gg or a $\cPq\bar\cPq$ initial state. However, other diagrams (not shown in the figure) also contribute, including $t$-channel diagrams, virtual coloron contributions in the $s$-channel, and a 4-point ggCC interaction.

The second class of models leading to the same final state involves axigluons A that arise from chiral color symmetry breaking $\text{SU}(3)_L\otimes \text{SU}(3)_R\to \text{SU}(3)_{L+R}$. In the benchmark model for axigluon pair production, two decay modes are considered. In the first mode (A1), the axigluon decays to scalar $\sigma$ and pseudoscalar $\tilde\pi$ color-octet states, each subsequently decaying to two gluons. Phenomenologically, this case is similar to coloron pair production, see Fig.~\ref{fig:diagram} (\cmsLeft). In the second mode (A2), the axigluon decays to a heavy color-triplet fermion Q in association with a light quark, and subsequently the heavy fermion decays to a standard model (SM) quark and a pseudogoldstone boson $\eta$, which is a light scalar particle with Higgs-like couplings, remaining from the left-right symmetry breaking via the Nambu--Goldstone mechanism~\cite{Nambu,Goldstone}. The $\eta$ boson then decays to a pair of fermions, where the fermions are the heaviest that are kinematically accessible, usually b quarks~\cite{AxigluonSigma}, as shown in Fig.~\ref{fig:diagram} (\cmsRight). In this case there is an additional \qqbar-induced axigluon pair production mode involving the $t$-channel exchange of Q.

\begin{figure*}[htb]
 \centering
 \includegraphics[width=0.48\textwidth]{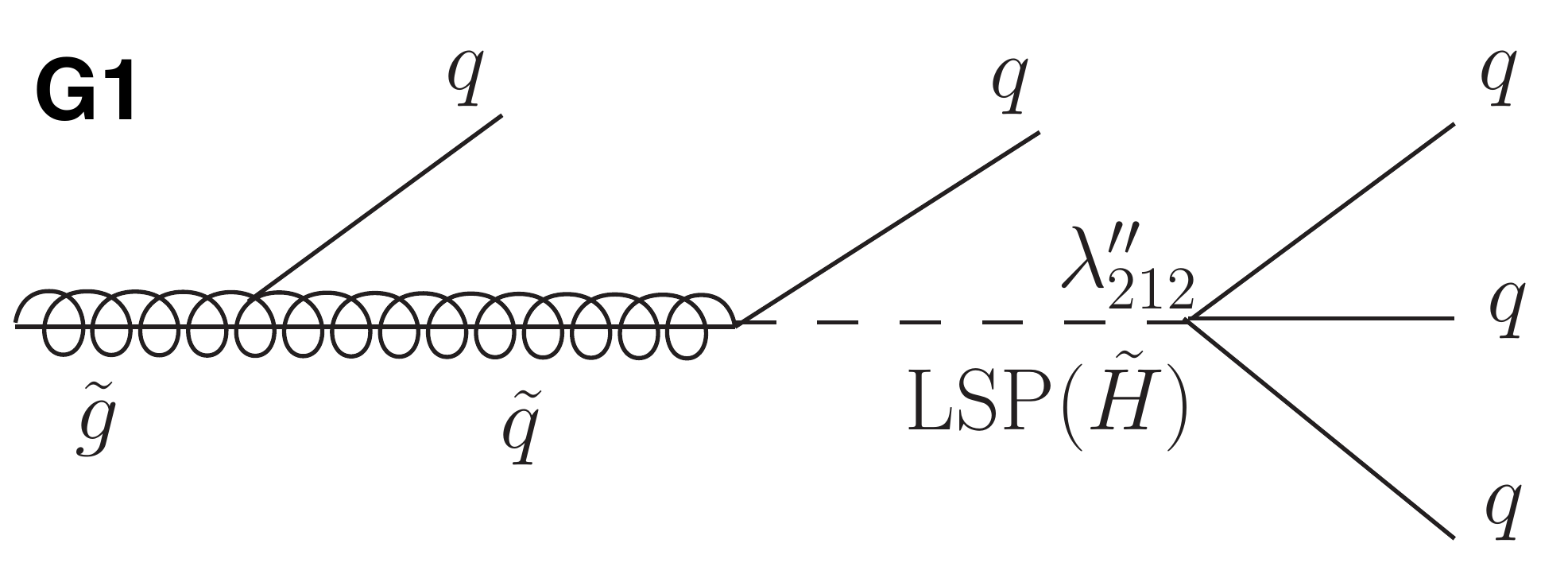}
 \includegraphics[width=0.48\textwidth]{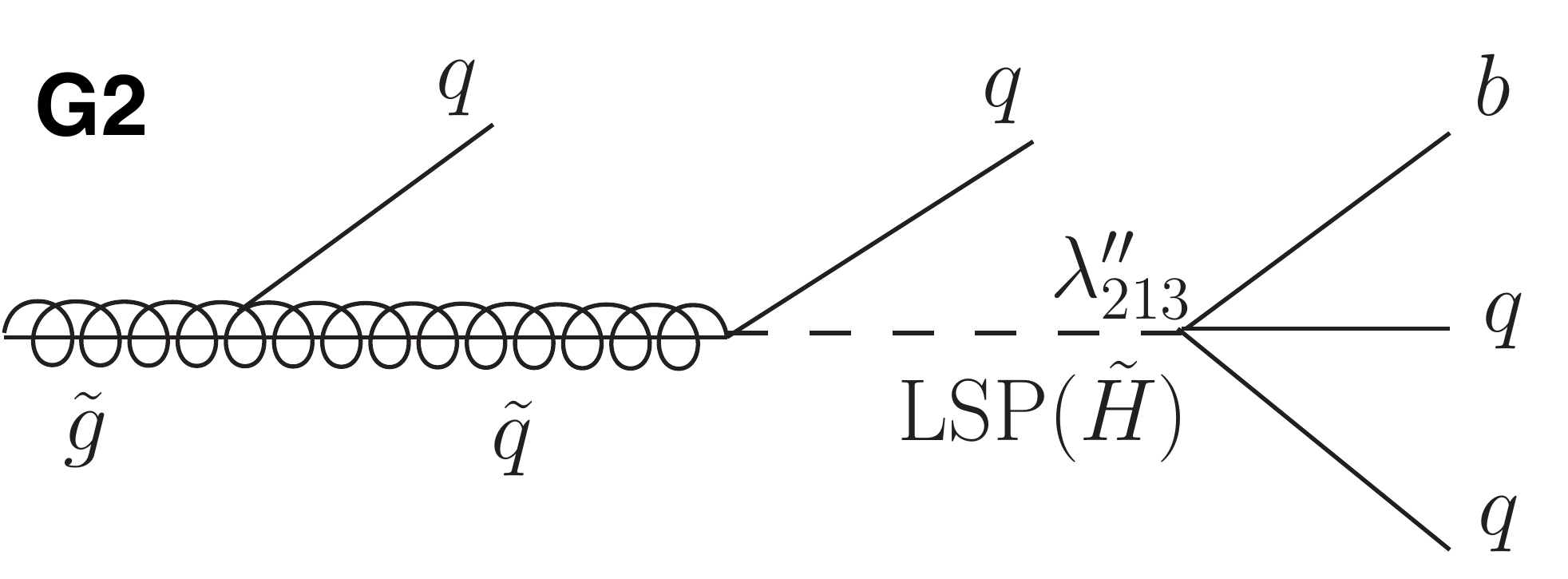}
 \includegraphics[width=0.48\textwidth]{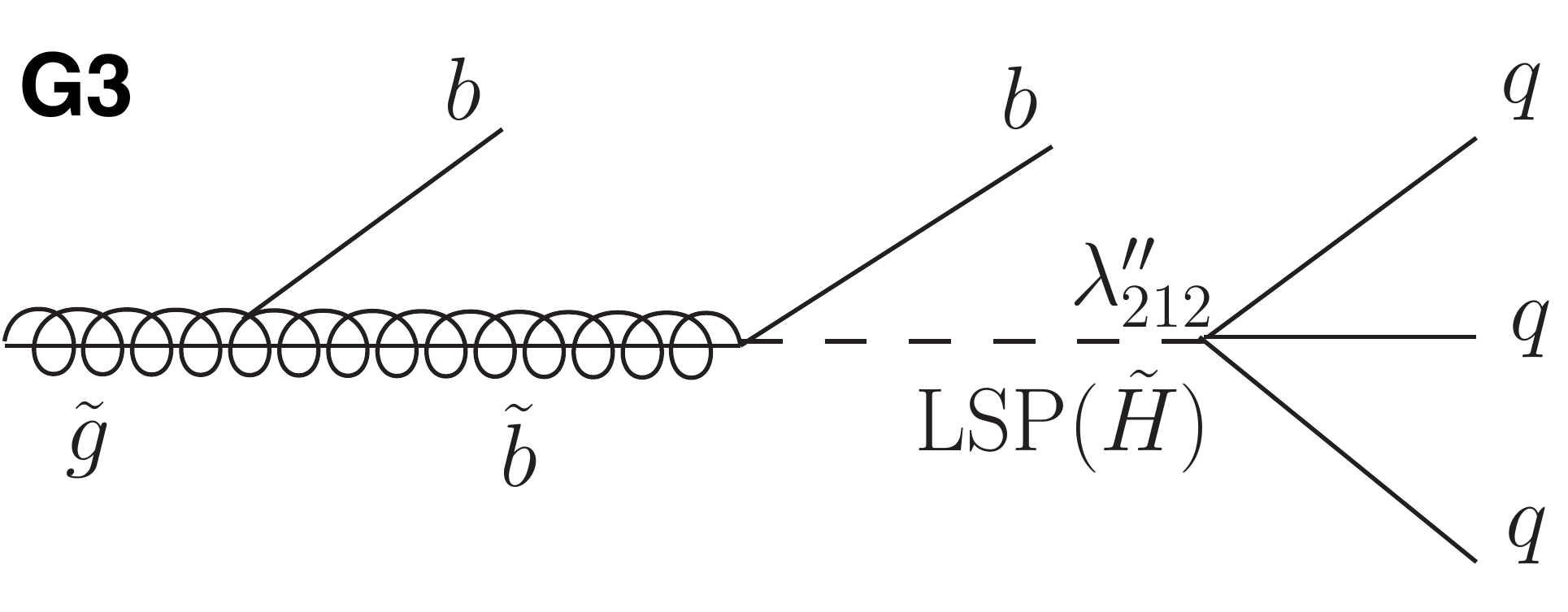}
 \includegraphics[width=0.48\textwidth]{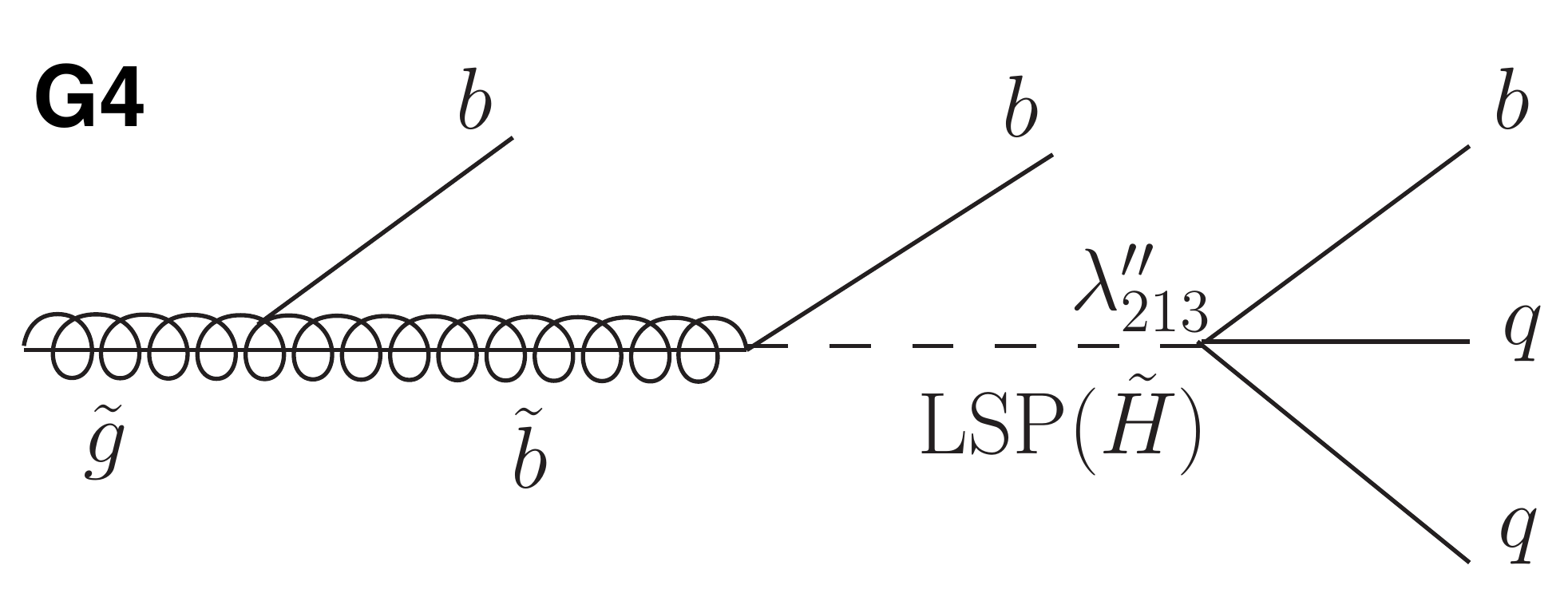}
 \caption{Gluino decay modes in the RPV SUSY scenario considered. Depending on the RPV coupling and the nature of the squark, zero (G1, top left), one (G2, top right), two (G3, bottom left), or three (G4, bottom right) b quarks can be present in each decay.}
\label{fig:diagramRPV}
\end{figure*}

The third class of models~\cite{RPVModel} involves an RPV SUSY scenario resulting in baryon number violation via the $\lambda''$ couplings (i.e., couplings that involve only quark superfields). We consider gluino (\PSg) pair production assuming that the second-generation squarks (\PSQ) are light, while the top squarks (\PSQt) are sufficiently heavy to prevent gluino decays involving top quarks. The decay chain is $\PSg \to  \cPq\PSQ$, followed by $\PSQ \to  \cPq\PSH$ and $\PSH \to \cPq\cPq\cPq$, where \PSH is the lightest supersymmetric particle (LSP) taken to be a higgsino. This process results in a 10-jet final state. Depending on the RPV coupling ($\lambda''_{212}$ or $\lambda''_{213}$) and on whether bottom squarks participate in the decay chain, either zero, one, two, or three of the quarks in this decay are b quarks, as shown in Fig.~\ref{fig:diagramRPV}. We refer to these gluino decay modes as G1, G2, G3, and G4, respectively.

\section{The CMS detector}

The central feature of the CMS apparatus is a superconducting solenoid of 6\unit{m} internal diameter, providing a magnetic field of 3.8\unit{T}. Within the solenoid volume are a silicon pixel and silicon strip tracker, a lead tungstate crystal electromagnetic calorimeter (ECAL), and a brass and scintillator hadron calorimeter (HCAL), each composed of a barrel and two endcap sections. Forward calorimeters extend the coverage in pseudorapidity $\eta$ provided by the barrel and endcap detectors. Muons are measured in gas-ionization detectors embedded in the steel flux-return yoke outside the solenoid.

The silicon tracker measures charged particles in the range $\abs{\eta}< 2.5$. For nonisolated particles  with transverse momentum $1 < \pt < 10\GeV$ and $\abs{\eta} < 1.4$, the track resolutions are typically 1.5\% in \pt and 25--90 (45--150)\mum in the transverse (longitudinal) impact parameter \cite{TRK}.

In the region $\abs{ \eta }< 1.74$, the HCAL cells have widths of 0.087 in pseudorapidity and azimuthal angle $\phi$. In the $\eta$-$\phi$ plane, and for $\abs{\eta}< 1.48$, the HCAL cells map on to 5 x 5 arrays of ECAL crystals to form calorimeter towers projecting radially outwards from close to the nominal interaction point. At larger values of $\abs{ \eta }$, the size of the towers increases. Within each tower, the energy deposits in ECAL and HCAL cells are summed to define the calorimeter tower energies, subsequently used to provide the energies and directions of hadronic jets. When combining information from the entire detector, the jet energy resolution amounts typically to 15\% at 10\GeV, 8\% at 100\GeV, and 4\% at 1\TeV, to be compared to about 40, 12, and 5\%, respectively, obtained when the ECAL and HCAL alone are used.

The first level of the CMS trigger system, composed of custom hardware processors, uses information from the calorimeters and muon detectors to select the most interesting events in a fixed time interval of less than 4\mus. A high-level trigger (HLT) processor farm further decreases the event rate from around 100\unit{kHz} to less than 1\unit{kHz}, before data storage.

A more detailed description of the CMS detector, together with a definition of the coordinate system used and the relevant kinematic variables, can be found in Ref.~\cite{CMS}.

\section{The Monte Carlo simulation}
\label{sec:MC}

While the dominant background in this analysis, stemming from QCD multijet production, is estimated using control samples in data, as detailed in Section~\ref{sec:background}, simulated background samples are used to qualify the background estimation methods and to ensure that other backgrounds are negligible. 

The QCD multijet background is simulated with the \MADGRAPH5 v5.1.3.30~\cite{bib:Madgraph,bib:Madgraph2,MadGraph} leading order (LO) Monte Carlo (MC) generator, interfaced with \PYTHIA 6.422~\cite{Sjostrand:2006za} for description of fragmentation and hadronization. Events are generated with the CTEQ6L~\cite{bib:CTEQ6L} parton distribution function (PDF) set. The underlying event is described using the \PYTHIA tune Z2*~\cite{Z2star1,Z2star2}. The generated events are processed through the full CMS detector simulation based on \GEANTfour~\cite{bib:GEANT4}. In addition, for an alternative description of the dominant background at high multiplicities, QCD multijet events are simulated using the \ALPGEN~\cite{ALPGEN} LO MC generator, with up to four additional outgoing partons in the matrix element calculations, also interfaced with \PYTHIA and \GEANTfour.

Three classes of signal models are simulated: pair-produced colorons; axigluons; and gluinos in an RPV SUSY scenario. Coloron production~\cite{MultiJetRes} is generated  using {\MADGRAPH4} v4.4.44~\cite{bib:Madgraph} followed by \PYTHIA for a specific production and decay mode (pair production of colorons, subsequently decaying to hyperpions and finally to gluons). The signal simulation is done for coloron masses $M_\mathrm{C}$ in the range from 0.4 to 1.5\TeV (in steps of 0.1\TeV), with a width $\Gamma_\mathrm{C}$ equal to 20\% of $M_\mathrm{C}$, and for a hyperpion mass equal to $M_\mathrm{C}$/3 (i.e. in the specific model of Ref.~\cite{Dicus2},  where the coloron decay to a pair of hyperpions is predicted to dominate). The detector simulation is performed using the CMS fast parametric simulation~\cite{bib:FASTSIM}. The results of the fast simulation have been cross-checked with results from the full simulation, for a few benchmark points, and the corresponding acceptances are found to agree within a few percent.

Axigluon pair production from gg and $\cPq\bar\cPq$ initial states through the gluon $s$-channel (Fig.~\ref{fig:diagram} \cmsLeft), axigluon $s$- and $t$-channel, heavy color-triplet fermion Q or SM quark $t$-channel exchange, and a 4-point interaction are simulated with \MADGRAPH5, followed by \PYTHIA and \GEANTfour. Two distinct decay topologies of pair-produced axigluons are considered in this analysis. The first topology is the decay of each axigluon to a pair of mass-degenerate color-octet scalar and pseudoscalar particles, each of which further decays to two gluons. The scalar particle mass $M_{\sigma}$ and the pseudoscalar particle mass $M_{\tilde\pi}$ are both chosen to be 1/4 or 1/3 of the axigluon mass $M_\mathrm{A}$, with $M_\mathrm{A}$ ranging from 0.1 to 1.5\TeV in steps of 0.1\TeV. The values of the mass ratio are chosen so that the decays of the axigluon to scalar and pseudoscalar particles dominate~\cite{AxigluonSigma}, and constraints imposed by 4-jet resonance searches~\cite{ATLAS-4j-1,ATLAS-4j-2,CMS-4j-1,CMS-4j-2} are avoided. Those searches would be sensitive to this model in the case of a light (pseudo)scalar when two gluon jets from its decay overlap and are reconstructed as a single jet. The width $\Gamma_\mathrm{A}$ of the axigluon is taken to be either 10 or 15\% of $M_\mathrm{A}$~\cite{AxigluonSigma}. The second topology is the decay of the axigluon to a heavy color-triplet fermion in association with a light quark (Fig.~\ref{fig:diagram} \cmsRight). The heavy quark subsequently decays to a quark and a light scalar ${\eta}$ that decays to a bottom quark-antiquark pair. This signal topology is simulated for $M_\mathrm{A}$ ranging from 0.4 to 1.5\TeV, in 0.1\TeV steps. The ratios are $M_{\rm Q}/M_\mathrm{A}$ = 2/3, $m_{\eta}/M_\mathrm{A} = 1/15$, and $\Gamma _\mathrm{A}/M_\mathrm{A} = 3.5$ or $10\%$, as recommended in Ref.~\cite{AxigluonSigma}. While for this choice of particle masses some merging of jets from a cascade decay of an axigluon does occur, the acceptance in the $\ge 8$ jet final state remains high (around 70\% for an axigluon mass of 0.7 TeV for the $H_T > 1.4$ TeV preselection used in the analysis).

The RPV SUSY gluino pair production and the decay chains~\cite{RPVModel} are simulated with \MADGRAPH5, interfaced with \PYTHIA and \GEANTfour. Gluino masses ${M_{\PSg}}$ from 0.5 to 1.5\TeV and squark masses ($M_{\PSQ}$ or $M_{\PSQb}$) from 0.1 to 0.9\TeV, in 0.1\TeV steps, are used. The higgsino mass is fixed to 3/4 of the relevant squark mass. The gluino pair production cross section is calculated with \textsc{nll-fast}~\cite{NLL-fast} at next-to-leading-order accuracy in $\alpha_S$ and with the resummation of soft-gluon emission at next-to-leading-logarithmic accuracy. The cross sections of pair-produced gluinos are identical in the four different decay scenarios considered. The values of the RPV couplings used in simulation are chosen to ensure prompt gluino decays.

All simulated samples include the effect of multiple proton-proton (pp) collisions per bunch crossing by superimposing minimum bias interactions with a multiplicity distribution matching that observed in data.

\section{Event selection and reconstruction}

The search described in this Letter utilizes a data sample of pp collisions at  $\sqrt{s} = 8\TeV$, collected with the CMS detector at the LHC in 2012 and corresponding to an integrated luminosity of 19.7\fbinv. The data were collected with a trigger based on jets reconstructed with the calorimeter-only information. At the HLT, the jets are clustered from the ECAL and HCAL energy deposits, using the anti-\kt clustering algorithm~\cite{bib:AKT} with a distance parameter of $R = 0.5$. Jet energies are corrected for the calorimeter response~~\cite{Chatrchyan:2011ds}. The trigger requires the scalar sum of \pt of all the HLT jets to exceed a threshold that was increased progressively from 550 to 750 \GeV to maintain a
constant trigger rate despite the increase in the instantaneous luminosity delivered by the LHC.

Offline, events are reconstructed using a particle-flow algorithm~\cite{PFT-09-001,PFT-10-001} that identifies each single particle (photon, electron, muon, charged hadron, and neutral hadron) with an optimized combination of all subdetector information. The energy of photons is directly obtained from the ECAL measurement corrected for zero-suppression effects. The energy of electrons is determined from a combination of the track momentum at the main interaction vertex, the corresponding ECAL cluster energy and the energy sum of all bremsstrahlung photons attached to the track. The energy of muons is obtained from the corresponding track momentum. The energy of charged hadrons is determined from a combination of the track momentum and the corresponding ECAL and HCAL energy, corrected for zero-suppression effects and for the response function of the calorimeters to hadronic showers. Finally, the energy of neutral hadrons is obtained from the corresponding corrected ECAL and HCAL energy.

For each event, hadronic jets are clustered from the particle-flow candidates, using the anti-\kt algorithm with a distance parameter of 0.5. The momentum of each jet is determined as the vectorial sum of all particle momenta in the jet and its magnitude is found in the simulation to be within 5 to 10\% of the true momentum at the particle level over the whole \pt spectrum and detector acceptance. Jet energy corrections are derived from simulation and are confirmed with in situ measurements using the energy balance of dijet and $\gamma$+jet events~\cite{Chatrchyan:2011ds}.

Events are further required to have at least one well-reconstructed~\cite{TRK} pp interaction vertex. In order to suppress jets due to rare, anomalous calorimeter signals, jet candidates are required to satisfy the following identification criteria: each jet should contain at least two particles, at least one of which is a charged hadron, and the jet energy fraction carried by neutral hadrons and photons should be less than 90\%. These criteria have an efficiency greater than $99\%$ per jet. Only events with at least 8 or 10 identified jets, depending on the search category, with $\pt>30\GeV$ and $\abs{\eta}<2.4$ are considered. Finally, the offline \HT variable, defined as a scalar sum of transverse momenta of all the jets passing the above requirements, must exceed $900\GeV$, in order to avoid any trigger bias.

For the signal channels with b quark jets in the final state, we require at least one jet to be b-tagged using the combined secondary vertex (CSV) algorithm~\cite{bib:btag,bib:BTagging}, which exploits information from tracks and secondary vertices to build a likelihood-based discriminator. This discriminator is then used to distinguish between jets originating from b quarks and those from c quarks, light-flavor quarks, and gluons. The  operating point of the CSV algorithm~\cite{bib:BTagging} used to tag b quark jets is defined by the minimum threshold on the discriminator at which the misidentification probability of light-parton jets is approximately 1\%; this working point corresponds to approximately 70\% tagging efficiency per b jet.
The effect of requiring more than one b-tagged jet has been investigated, particularly for the decay channels containing four or six b quark jets in the final state. It is found that this does not improve the search sensitivity, since the uncertainty in the estimation of background from data (as described in Section~\ref{sec:background}) becomes large. Therefore we stay with the requirement of a minimum of only one b-tagged jet for all signal channels with b quarks, independent of their multiplicity.

In addition to these selection criteria, for the channels without b quark jets in the final state we employ a global event shape variable, sphericity ($S$)~\cite{Bjorken}. This variable is based on the three eigenvalues $Q_1\ge Q_2 \ge Q_3$ of the tensor $S^{\alpha\beta}$ in the momentum space:
$S^{\alpha\beta} = { \sum_{i}p_i^{\alpha}p_i^{\beta}/\sum_i p_i^2}$, where indices $\alpha$ and $\beta$ run over the three spatial coordinates and $p_i$ is the momentum of jet $i$. The sphericity is defined as $S = \frac{3}{2}(Q_{2}+Q_{3})$. Events with $S \approx 1$ are more spherically symmetric, whereas events with $S \approx 0$ are more linear, looking like a pair of back-to-back jets. The signal events are more spherical than the background events, which are largely characterized by the back-to-back topology of the jets from the hard-scattering 2$\to$2 parton processes. This shape variable was previously used in the search for light- and heavy-flavor three-jet resonances~\cite{CMS-6j-3}, to separate the signal from the QCD background. The optimum selection on the sphericity value was determined by maximizing the expected signal significance, while keeping the selection soft enough to maintain the invariance of the \HT distribution with respect to the jet multiplicity (explained in detail in Section~\ref{sec:background}), and corresponds to $S > 0.1$. Using the sphericity variable increases the expected signal significance by 10\%.

Table~\ref{table:sr} summarizes the four signal regions (SR1--SR4) used in the analysis and the models probed by each signal region, as discussed in more detail in Section~\ref{sec:results}.

\begin{table*}[htb]
  \centering
  \topcaption{Definition of signal regions used in the analysis, and models probed by each signal region.}
  \begin{tabular}{cll}
  Signal region & \multicolumn{1}{c}{Selection} & \multicolumn{1}{c}{Models probed}\\
  \hline
  SR1 & ${\ge}8$ jets (\PT $> 30$\GeV, $\abs{\eta} < 2.4$), $S > 0.1$ & Colorons, A1, low-mass G1 \\
  SR2 & ${\ge}8$ jets (\PT $> 30$\GeV, $\abs{\eta} < 2.4$), ${\ge}1$ b-tagged & A2, low-mass G2, G3, G4 \\
  SR3 & ${\ge}10$ jets (\PT $> 30$\GeV, $\abs{\eta} < 2.4$), $S > 0.1$ & High-mass G1 \\
  SR4 & ${\ge}10$ jets (\PT $> 30$\GeV, $\abs{\eta} < 2.4$), ${\ge}1$ b-tagged & High-mass G2, G3, G4 \\
  \end{tabular}
  \label{table:sr}
\end{table*}

\section{Background estimation}
\label{sec:background}

Because of the large combinatorial background in the multijet final states, we do not employ any mass variables in the analysis. The results are based on simultaneous counting experiments in the low-\HT control region, dominated by the background, and in the signal region above a certain \HT cutoff, optimized for each signal point, as discussed below. The main background in this analysis is QCD multijet production, which is estimated directly from the observed data using the \HT multiplicity invariance method ~\cite{CMSBH1,CMSBH2,CMSBH3}, extensively used for black hole searches in CMS. The method is based on an empirical observation that the shape of the \HT spectrum is independent of the jet multiplicity, which was thoroughly checked by using various MC samples (\PYTHIA, \ALPGEN, and \MADGRAPH5), as well as collision data. The invariance stems from the fact that, in a generic QCD event, the \HT distribution is approximately determined by the 2$\to$2 hard-scattering processes. Any further splitting of jets as a result of FSR conserves the \HT, as long as both the hard-scattering and FSR jets still pass the \pt selection. The initial-state radiation, which is predominantly a forward process, does potentially change the \HT value, but this turns out to be a subdominant effect and does not spoil the observed invariance. For the purpose of this analysis, we have also confirmed that the \HT invariance is preserved with the relatively soft sphericity  selection we use in the non-b-tagged analysis; it is also preserved when at least one b-tagged jet is required. The \HT invariance method consists of fitting the shape of the \HT distribution at lower jet multiplicities and then using this shape to describe the background for the high-multiplicity signal selection.

For the searches in inclusive 8- and 10-jet final states, fits to five analytic functions are performed for \HT distributions in data control samples with an exclusive jet multiplicity requirement between 4 and 7, with and without the requirement of at least one b-tagged jet. The potential signal contamination is shown to be small in these lower-multiplicity samples. The fit functions are similar to the one used in the searches for dijet resonances~\cite{CMS1,CMS2,CMS3,CMS4,CMS5,CMS6,CMS7,CMS8,CMS9,CMS10,CMS11} or to those used in searches for microscopic black holes~\cite{CMSBH1,CMSBH2,CMSBH3} in CMS:
\begin{itemize}
\item $f_1 = \frac{P_0 (1 + x)^{P_1}}{x^{P_2 + P_3 \ln(x)}}$, 
\item $f_2 = \frac{P_0}{(P_1 + P_2 x + x^2)^{P_3}}$,
\item $f_3 = \frac{P_0}{(P_1 + x)^{P_2}}$, 
\item $f_4 = \frac{P_0 (1 + x)^{P_1}}{x^{P_2 \ln(x)}}$,
\item $f_5 = \frac{P_0 (1 - x)^{P_1}}{x^{P_2 + P_3 \ln(x)}}$,
\end{itemize}
where $x = \HT/\sqrt{s}$, and $P_i$ are the free parameters of the fit.

All five fits are consistent with each other and have good fit probability. We pick the function that fits best a particular control sample used to predict the background in one of the four search regions. The other fit functions and control samples are used to determine the background uncertainties, as described in Section~\ref{sec:syst}. The $f_3$ function is used to fit to the 4-jet \HT spectrum without any sphericity requirement as the background template for SR1 (coloron and axigluon A1 searches). The same function fit to the 4-jet \HT spectrum with at least one b-tagged jet requirement is used to predict the background in the SR2 (A2 and low-mass G2, G3, and G4 searches). The $f_4$ function fit to the 4-jet \HT spectrum without any sphericity requirement is used as the background template for the RPV gluino G1 search at high masses (SR3), while for the high-mass G2, G3, and G4 scenarios (SR4) the $f_2$ function fit to the 4-jet \HT spectrum with at least one b-tagged jet is used as the background template. The fit range is chosen to be 1.5--2.0\TeV to predict the background shape for the ${\ge}8$ jet search regions SR1 and SR2, and 2.0--3.0\TeV for the ${\ge}10$ jet search regions SR3 and SR4. The fit ranges were chosen to be above the turn-on of the multiplicity invariance in the corresponding search regions and to allow for an adequate statistical precision of the fits. The effect of the fit range variation was shown to be small compared with the other uncertainties in the background prediction, detailed in Section~\ref{sec:syst}.

For low-mass resonances the signal contamination can be significant, even in the low-\HT range of the spectrum; therefore, the signal extraction (or limit setting) procedure has been generalized to take potential contamination into account. For each signal region, we define the control bin and the signal bin in the \HT distribution to be used in the simultaneous counting experiment to extract the background normalization and the potential signal. The control bin is chosen to be 1.4--1.7\TeV for SR1 and SR2, and 1.9--2.1\TeV for SR3 and SR4, where the lower boundary is chosen to avoid the multiplicity-related turn-on effects, and the upper boundary is chosen low enough to minimize signal contamination. In the limit of no signal contamination, the control bin essentially becomes the background normalization region. The signal bin is defined by the requirement of \HT to exceed a certain threshold, which is determined for each resonance mass in each model to maximize the signal significance. For the case where the background expectation exceeds 20 events, the Gaussian significance $S/\sqrt{B}$ is used, where $S$ and $B$ are signal and background expectations. For high-mass resonances, where the optimal \HT thresholds are sufficiently high to have a small number of expected background events, the $Z_{Bi}$ criterion~\cite{zbi} is used instead. The $Z_{Bi}$ statistic is a measure of equivalent Gaussian signal significance obtained by considering the binomial probability of the events in data being distributed at least as signal-like as observed, under the assumption of the background-only hypothesis.

\begin{figure*}[htbp]
  \centering
  \includegraphics[width=0.48\textwidth]{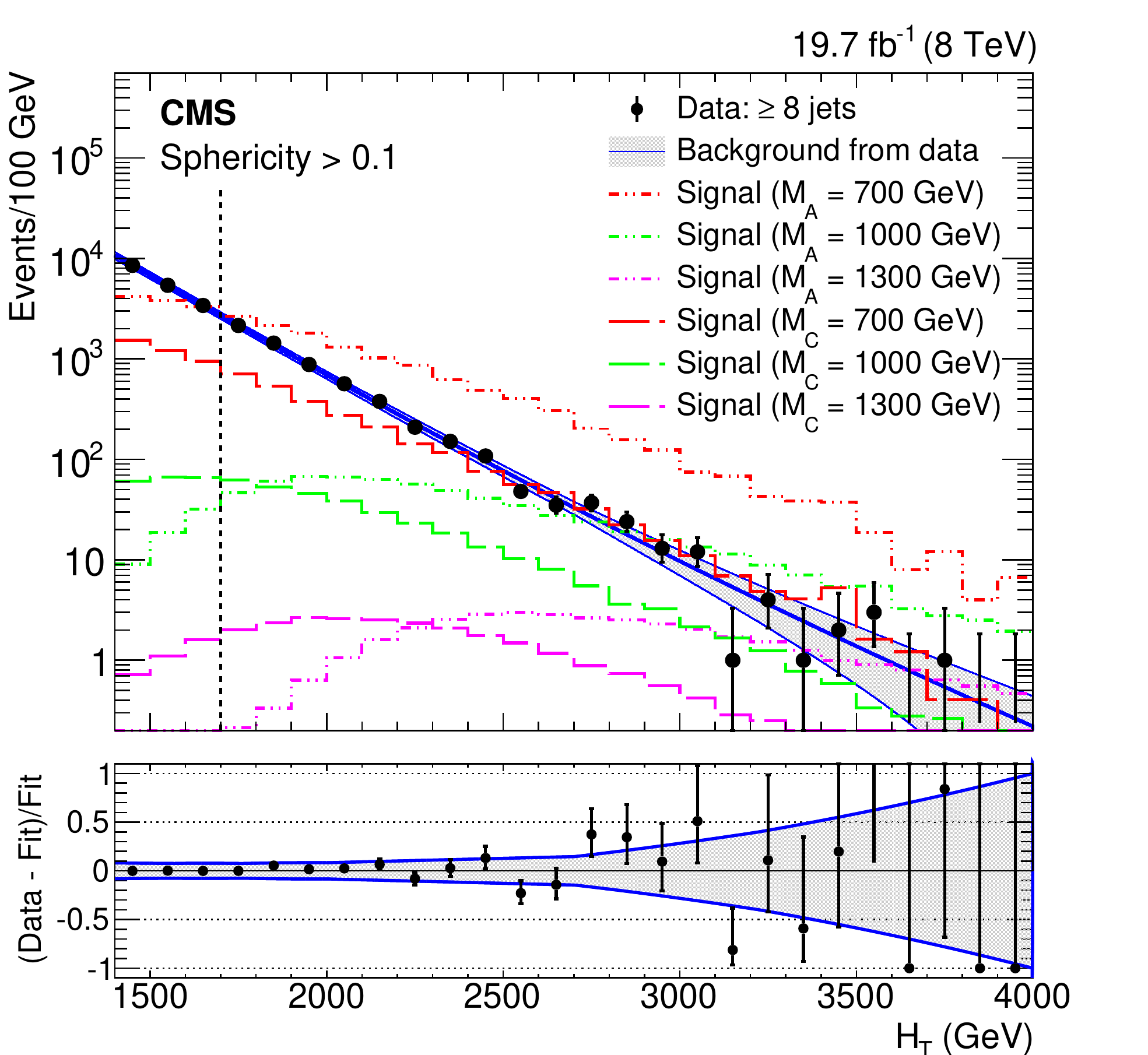}
  \includegraphics[width=0.48\textwidth]{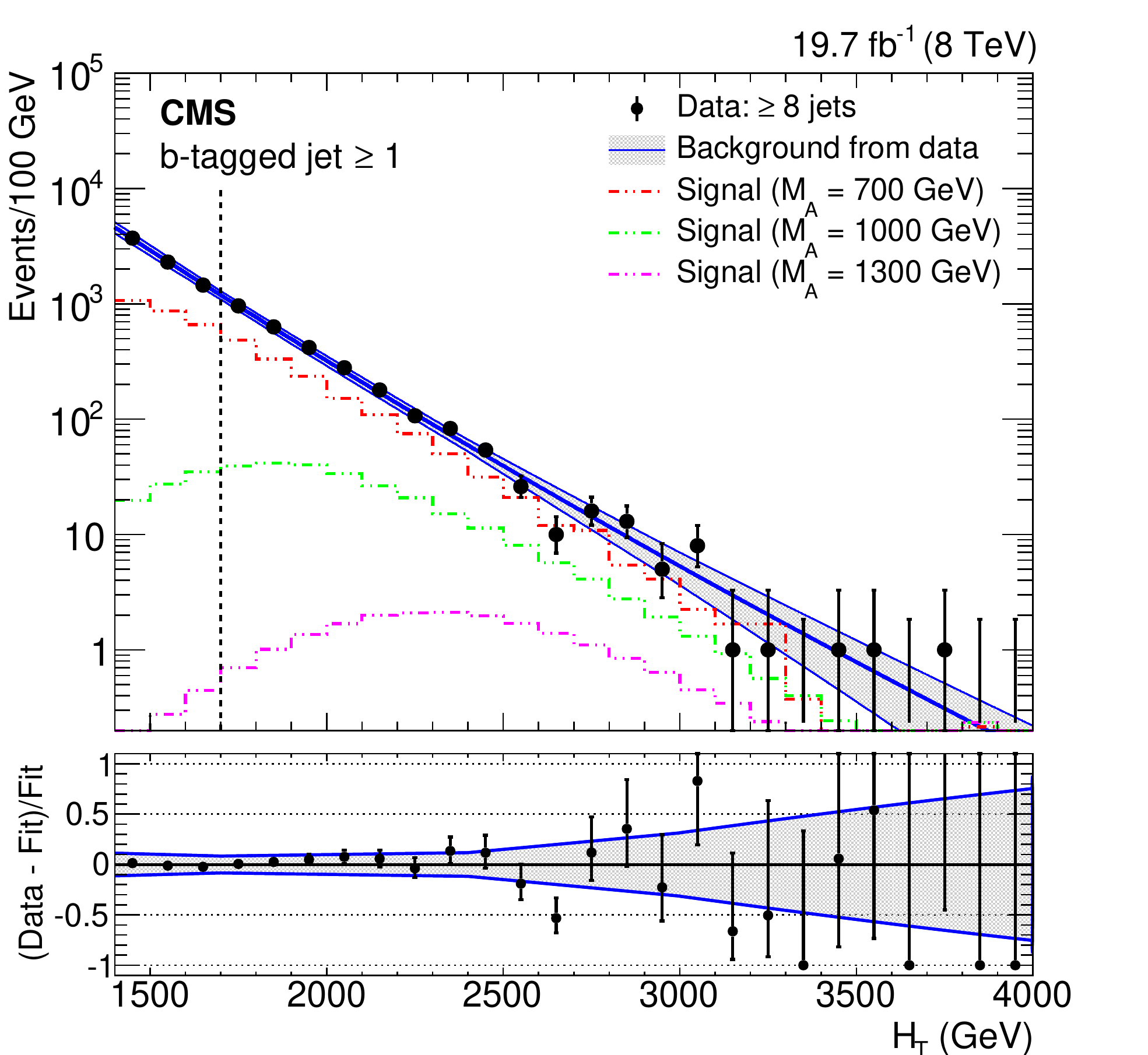}
\includegraphics[width=0.48\textwidth]{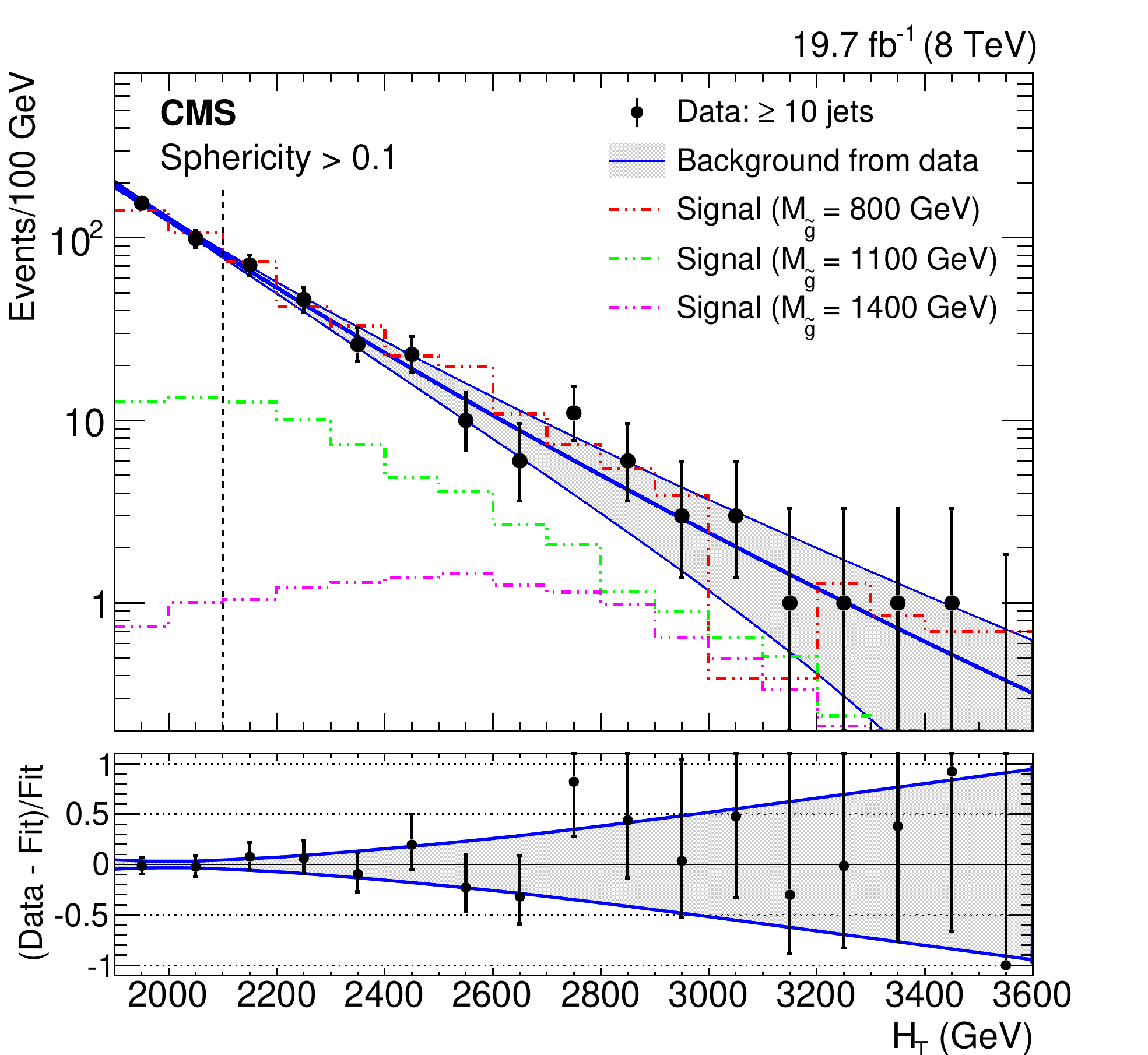}
\includegraphics[width=0.48\textwidth]{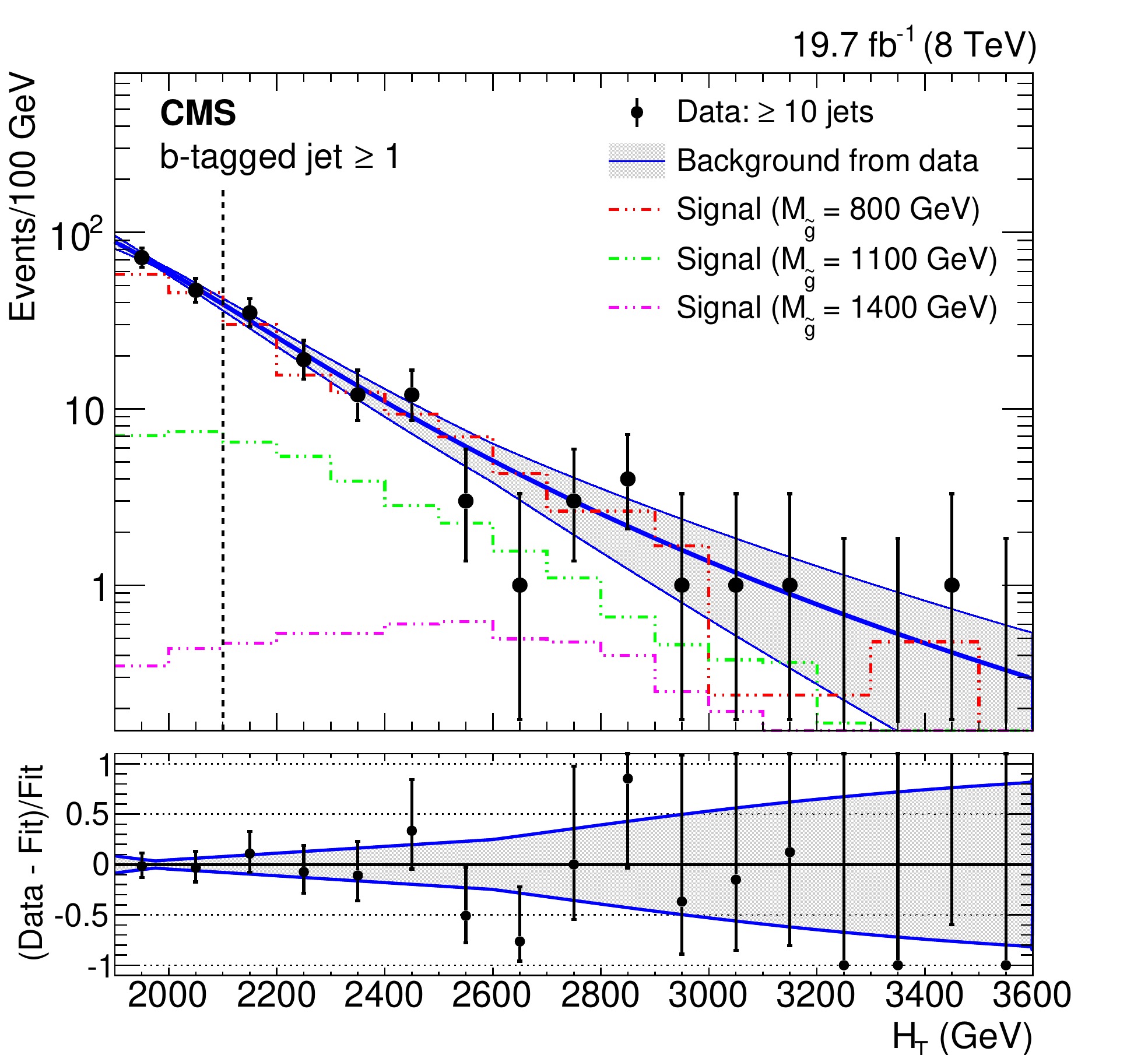}
 \caption{The $\HT$ distributions in data (points), background estimated from data (blue thick solid line in the upper panels) with its uncertainty (gray shaded band), and representative signal model predictions (histograms). Top left: 8 or more jets, no b tagging requirement, with the coloron ($\Gamma_\mathrm{C}/M_\mathrm{C} = 20\%$, $M_{\tilde\pi} = M_\mathrm{C}/3$)  and axigluon A1 ($\Gamma_\mathrm{A}/M_\mathrm{A} = 15\%$, $M_{\sigma/\tilde\pi} = M_\mathrm{A}/3$) signals overlaid. Top right: 8 or more jets, including one or more b-tagged jets, with the A2 ($\Gamma_\mathrm{A}/M_\mathrm{A} = 15\%$, $M_{\sigma/\tilde\pi} = M_\mathrm{A}/3$) axigluon signal points overlaid. Bottom left (right): 10 or more jets without b tagging requirement (with one or more b-tagged jets), with RPV SUSY gluino G1 (G2) signals with a squark mass of 400 GeV overlaid. The lower panels show the distribution of the quantity (Data $-$ Fit)/Fit. The error bars on the plotted values indicate the statistical uncertainty associated with the data, and the shaded band indicates the systematic uncertainty. Dashed vertical lines indicate the upper boundary of the control bins.}
  \label{fig:FinalBHDistribution}
\end{figure*}

The \HT distributions in data, with background estimated from control samples in data, and relevant signal model predictions are shown in Fig.~\ref{fig:FinalBHDistribution} for the four signal regions. In all four cases good agreement between the data and the background predictions is observed.

\section{Systematic uncertainties}
\label{sec:syst}

Several sources of systematic uncertainties in signal acceptance and background prediction have been considered in the analysis.

Two main sources of the background uncertainty are the choice of the fit function (shape uncertainty) and the normalization uncertainty. In order to estimate the shape uncertainty, we use the envelope of the fits with five template functions to the \HT spectrum in the 4- to 7-jet control samples in data. This uncertainty ranges from approximately 3\% for $\HT = 1.4$\TeV to 140\% for $\HT = 4$\TeV. The background normalization uncertainty is statistical in nature, because of the limited number of data events in the normalization region. It varies between 2 and 10\%, depending on the signal region. The uncertainty related to the assumption of \HT invariance is included in the shape uncertainty, as the fit function envelope contains fits to several  exclusive low-multiplicity distributions, which allows one to gauge the degree to which the \HT invariance may be violated.

Systematic uncertainties in the signal acceptance are related to the choice of PDFs and the uncertainty in the jet energy scale. The PDF uncertainty is estimated using the PDF4LHC prescription~\cite{PDF4LHC1,PDF4LHC2}, based on the CT10~\cite{CT10}, MSTW2008~\cite{MSTW}, and NNPDF2.1~\cite{NNPDF} sets and found to be 3\%. The uncertainty due to the jet energy scale is estimated by varying the latter up and down by one standard deviation and estimating the effect on the signal acceptance after all the selections. This procedure gives a 5\% uncertainty in the acceptance due to the uncertainty in the jet energy scale. In the case of signal models with b quark jets in the final state, there is an additional systematic uncertainty due to the uncertainty in the b tagging efficiency scale factors that account for the difference between the b tagging performance in data and in simulation~\cite{bib:BTagging}. This uncertainty is taken into account by varying the scale factor values up and down by one standard deviation and estimating the effect on the signal acceptance. The resulting uncertainty ranges between 2 and 5\%, depending on \HT. Finally, a 2.6\% uncertainty in the integrated luminosity~\cite{lumi} is also applied to the signal yield.

The systematic uncertainties in the signal and background yields are summarized in Table~\ref{tb:BHSystematics}.

\begin{table*}[htb]
  \centering
  \topcaption{Summary of systematic uncertainties in the signal yields and background yields}
    \label{tb:BHSystematics}
  \begin{tabular}{lcc}
  Uncertainty source & Signal uncertainty & Background uncertainty\\
  \hline
  Jet energy scale      &      5\%         &       \NA      \\

  PDF                   &      3\%         &       \NA      \\

  b tagging scale factor &      2--5\%         &   \NA      \\

  Integrated luminosity            &         2.6\%          &   \NA \\

  Background shape         &         \NA          &   3--140\%\\

  Background normalization            &         \NA           &   2--10\%\\

  \end{tabular}
\end{table*}

\section{Results}
\label{sec:results}

We construct the following likelihood to describe the results of two simultaneous counting experiments with $n_\mathrm{C}$ ($n_\mathrm{S}$) events observed in the control (signal) bin:
\ifthenelse{\boolean{cms@external}}{
\begin{multline}
\label{eq:likelihood}
L(\mu, k, \vec{\theta}) = \frac{\re^{-(\mu S_\mathrm{C}(\vec{\theta}) + k B_\mathrm{C}(\vec{\theta}))} \left(\mu S_\mathrm{C}(\vec{\theta}) + kB_\mathrm{C}(\vec{\theta})  \right)^{n_{C}} }{n_{C}!} \\
\times \frac{\re^{-(\mu S_\mathrm{S}(\vec{\theta}) + k B_\mathrm{S}(\vec{\theta}))} \left(\mu S_\mathrm{S}(\vec{\theta}) + kB_\mathrm{S}(\vec{\theta})  \right)^{n_{S}} }{n_{S}!}.
\end{multline}
}{
\begin{equation}
\label{eq:likelihood}
L(\mu, k, \vec{\theta}) = \frac{\re^{-(\mu S_\mathrm{C}(\vec{\theta}) + k B_\mathrm{C}(\vec{\theta}))} \left(\mu S_\mathrm{C}(\vec{\theta}) + kB_\mathrm{C}(\vec{\theta})  \right)^{n_{C}} }{n_{C}!} \frac{\re^{-(\mu S_\mathrm{S}(\vec{\theta}) + k B_\mathrm{S}(\vec{\theta}))} \left(\mu S_\mathrm{S}(\vec{\theta}) + kB_\mathrm{S}(\vec{\theta})  \right)^{n_{S}} }{n_{S}!}.
\end{equation}
}
In this expression, the parameter $\mu$ is the scale factor for the signal (``signal strength"), $k$ is the normalization factor of the background template, $S_\mathrm{C}$ and $B_\mathrm{C}$ ($S_\mathrm{S}$ and $B_\mathrm{S}$) are the expected signal and background yields in the control (signal) bin, and $\vec\theta$ is the vector of nuisance parameters.

By maximizing the likelihood of Eq. (\ref{eq:likelihood}) with $\vec\theta$ profiled as log-normal nuisance parameters, we extract the best fit value and 95\% confidence level (CL) upper limit for the signal strength, which we convert to a limit on the cross section times the branching fraction for the multijet channel of the specific signal. For limit setting, we use the asymptotic approximation~\cite{bib:asCLs} of the CL$_s$ method~\cite{CLS1,CLS2}. When comparing the result with theoretical cross sections to extract mass limits, we assume this branching fraction is equal to one, i.e. that the decay proceeds exclusively in the specific mode we probe, with the exception of the coloron model, where the branching fraction is calculated following Ref.~\cite{Dicus2}, and typically exceeds 95\%.

Despite the potentially sizable signal contribution in the control bins, the sensitivity to the low-mass signals is still sufficiently high because of the large production cross section and the fact that the signal and background have different shapes; therefore, the fractional contributions of the signal in the control and signal bins are different. For the low-mass gluino ($M_\PSg < 1.1$\TeV), the sensitivity in the SR1/SR2 is actually higher than in the SR3/SR4, despite the presence of 10 jets in the final state resulting from the gluino decays, because the control bins in higher-multiplicity search regions SR3/SR4 suffer from potentially large signal contamination. The contamination  is significantly less pronounced in the control bins of lower-multiplicity regions SR1/SR2, which also correspond to lower values of \HT and are thus better separated from the signal.  Nevertheless, the minimum \HT requirement and the position of the control bins do not allow us to probe masses below 0.6\TeV, so all the results are quoted for signals with the masses above this threshold.

\begin{figure}[tbh]
  \centering
  \includegraphics[width=0.49\textwidth]{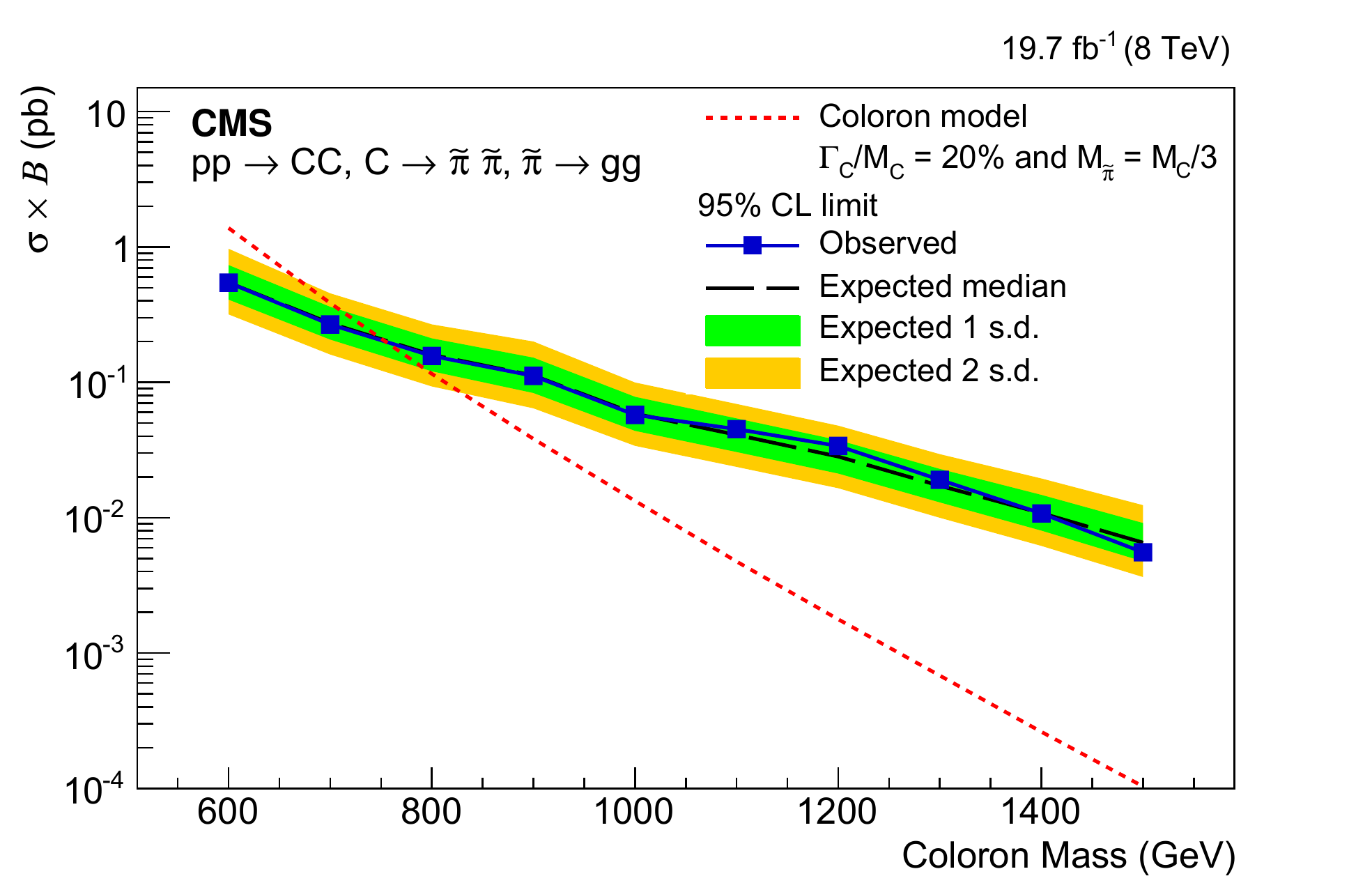}
 \caption{Upper limits at 95\% CL on the signal cross section times branching fraction, as a function of coloron mass $M_\mathrm{C}$, assuming a width of 20\% and a hyperpion mass $M_{\tilde\pi}$ equal to $M_\mathrm{C}/3$. The observed cross section limits (points) are compared with the expected limit (dashed line) and the one and two standard deviation uncertainty bands. The cross section for coloron pair production (dashed red line) is also shown.}
  \label{fig:LimitsColoronwidth20}
\end{figure}

The upper limits on the signal cross section times branching fraction at 95\% CL for the coloron model are shown in Fig.~\ref{fig:LimitsColoronwidth20}. By comparing the limits with the theoretical cross section times the branching fraction for coloron pair production, we exclude colorons with masses, $M_\mathrm{C}$, from 0.6 to 0.75\TeV for a hyperpion mass equal to $M_\mathrm{C}/3$.

\begin{figure*}[hbt]
  \centering
  \includegraphics[width=0.49\textwidth]{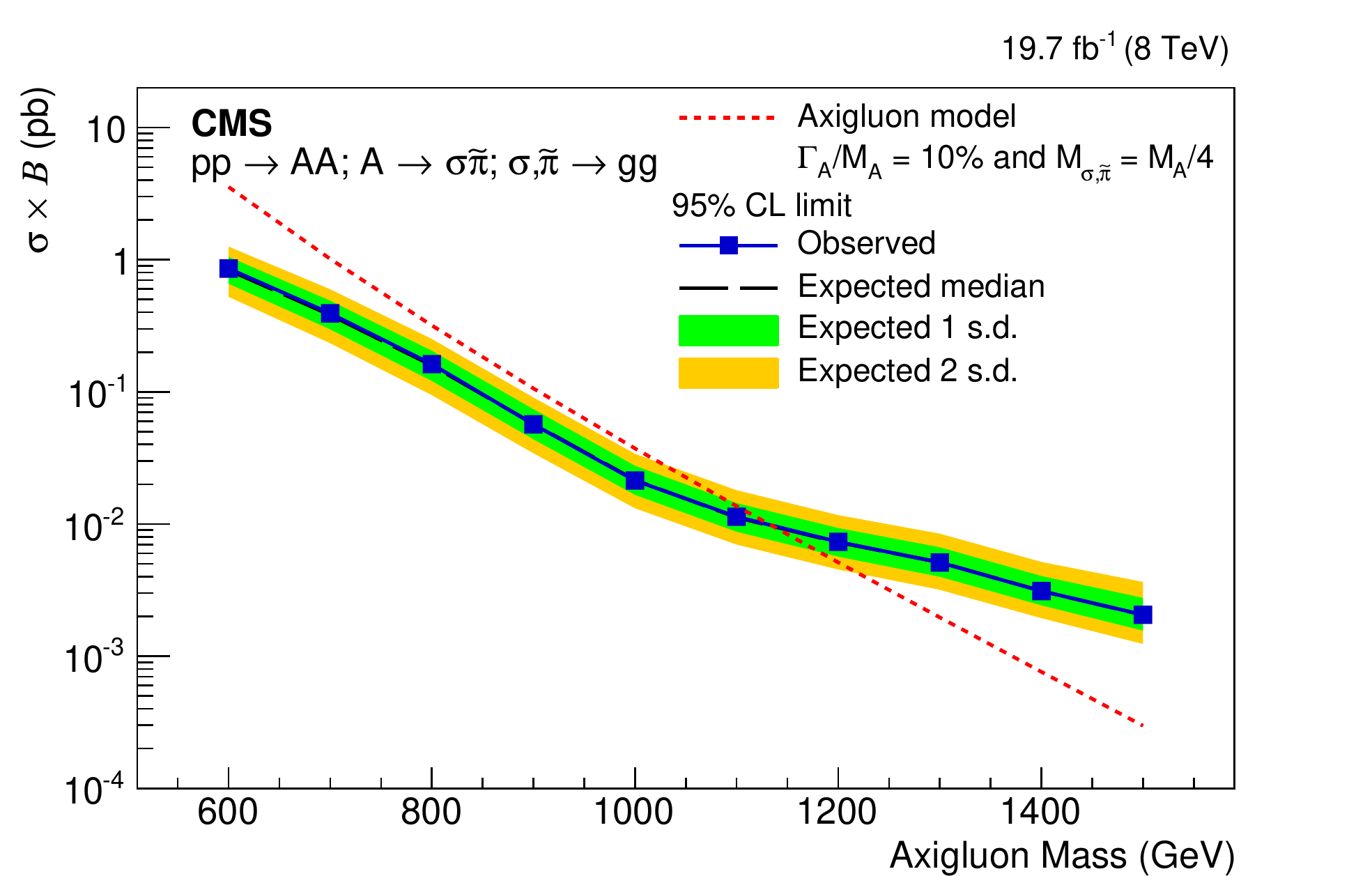}
  \includegraphics[width=0.49\textwidth]{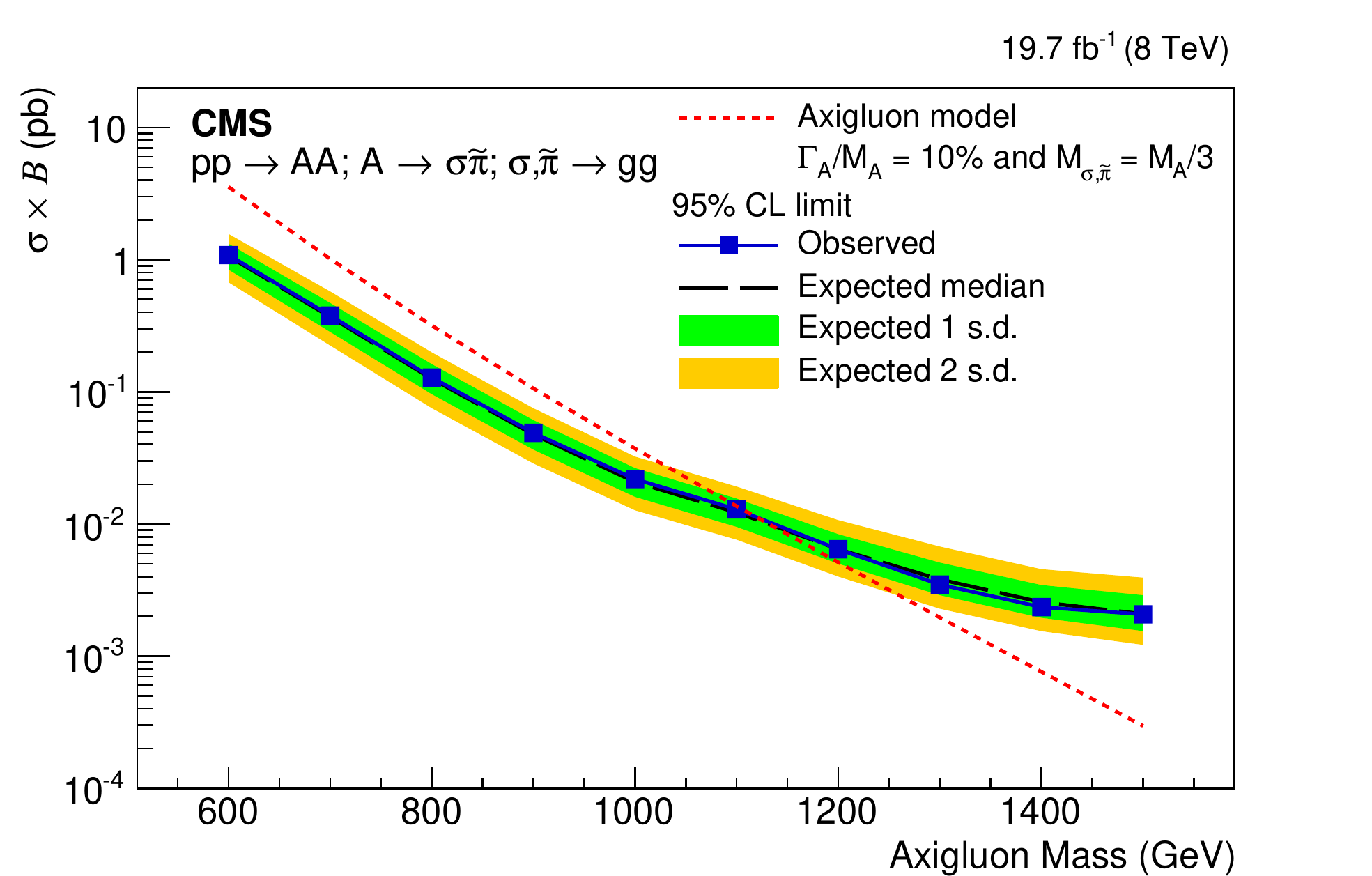}
  \includegraphics[width=0.49\textwidth]{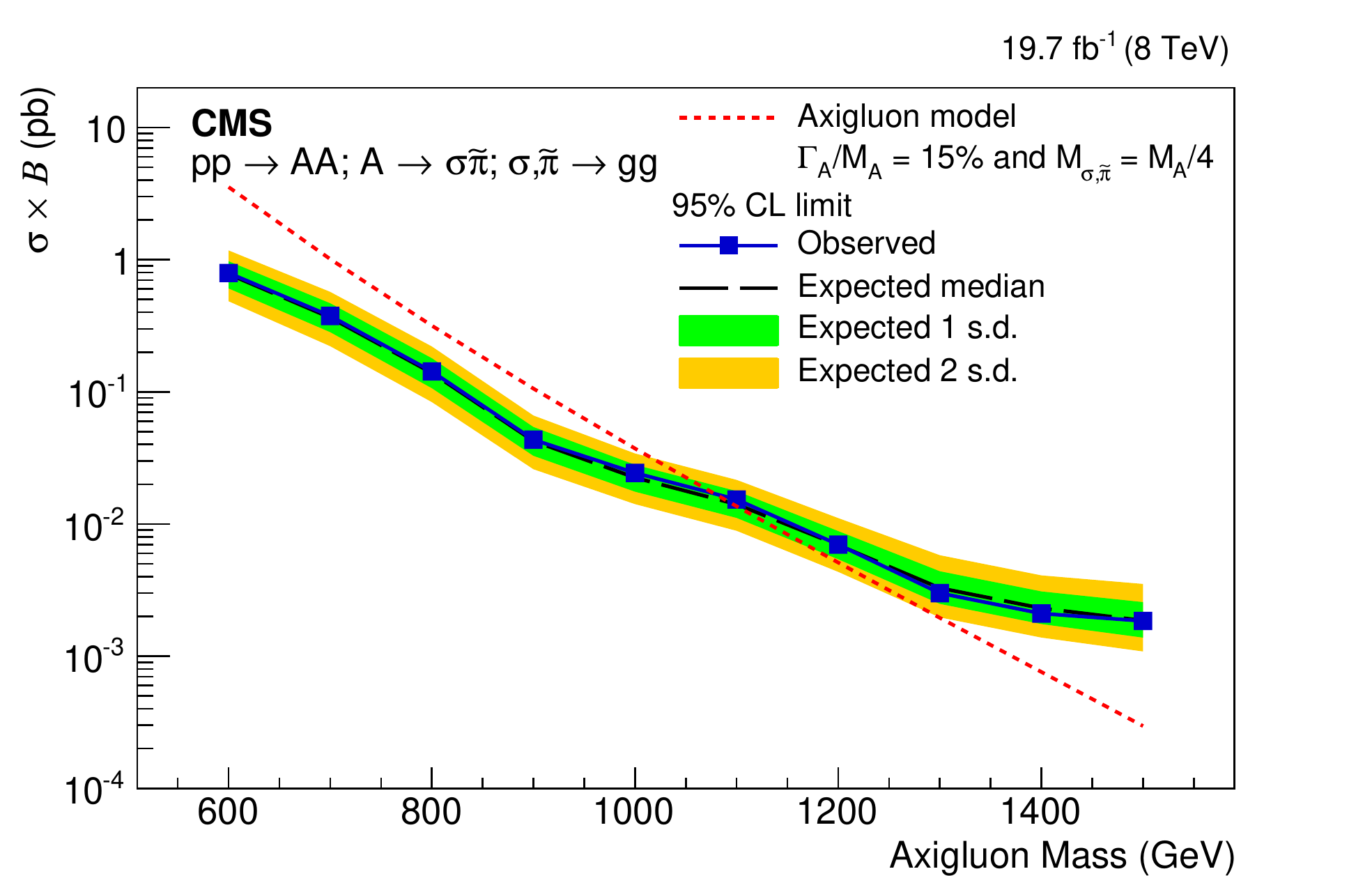}
  \includegraphics[width=0.49\textwidth]{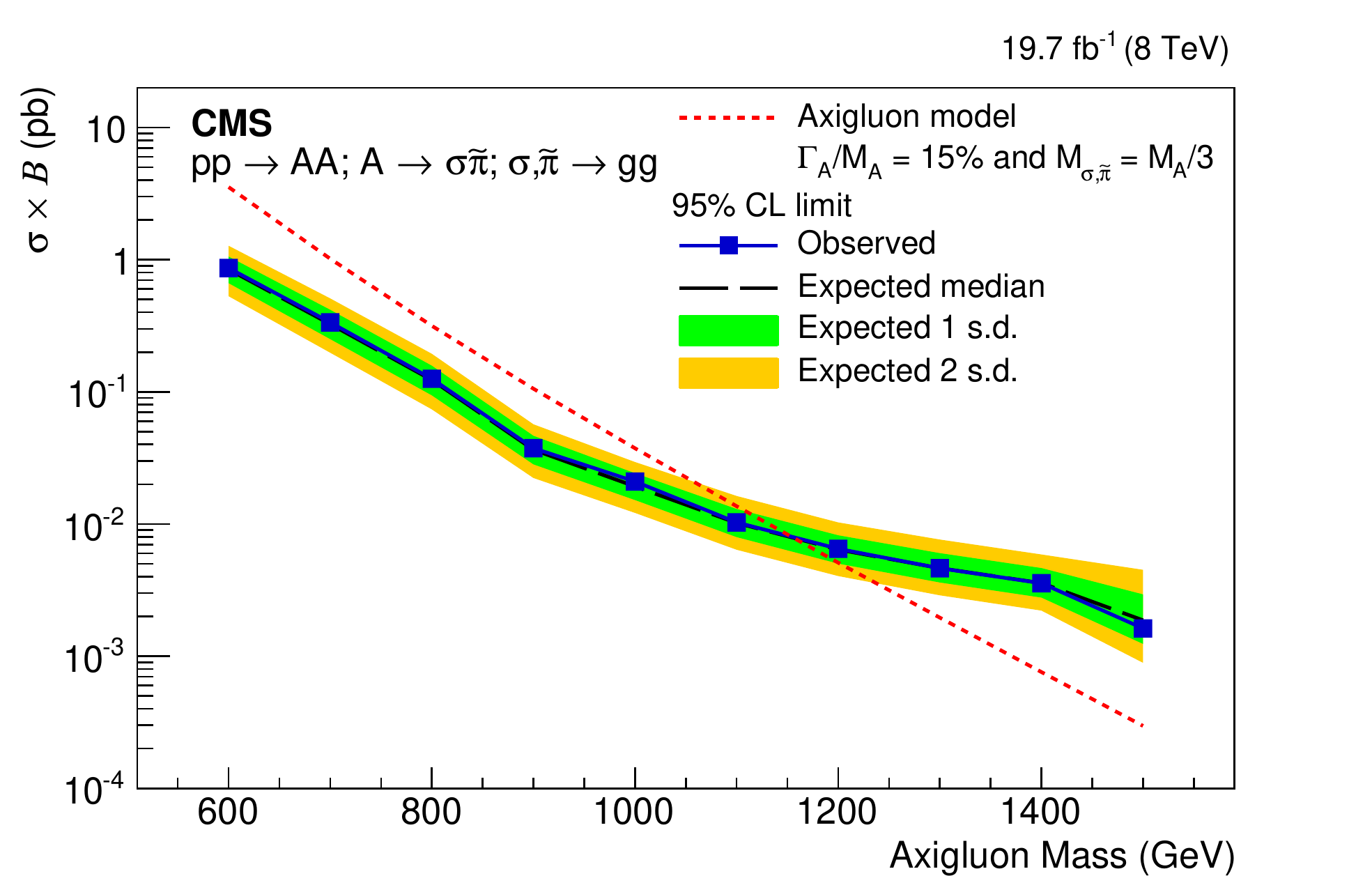}
  \caption{Upper limits at $95\%$ CL on signal cross section times branching fraction, as a function of axigluon mass $M_\mathrm{A}$, assuming a width of $10\%$ (top row) or 15\% (bottom row) and a decay according to the A1 model. Left: for equal scalar and pseudoscalar particle masses ($M_{\sigma} = M_{\tilde\pi} = M_\mathrm{A}/4$); right: for $M_{\sigma} = M_{\tilde\pi} = M_\mathrm{A}/3$. The observed cross section limits (points) are compared with the expected limit (dashed line) and the one and two standard deviation uncertainty bands. The cross section for axigluon pair production (dashed red line) is also shown.}
  \label{fig:LimitsAxiwidth10-15}
\end{figure*}

\begin{figure}[tbh]
  \centering
  \includegraphics[width=0.49\textwidth]{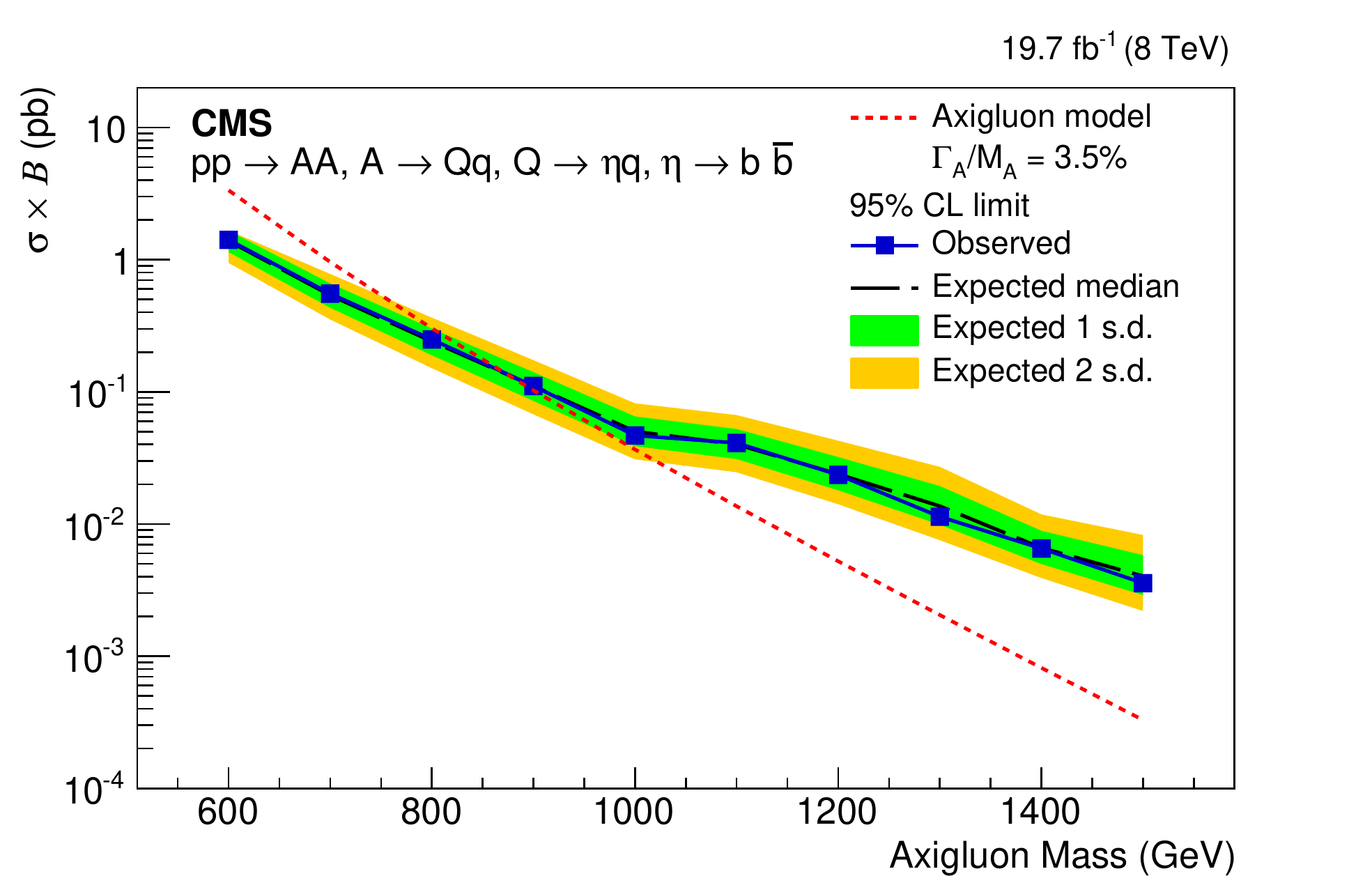}
  \includegraphics[width=0.49\textwidth]{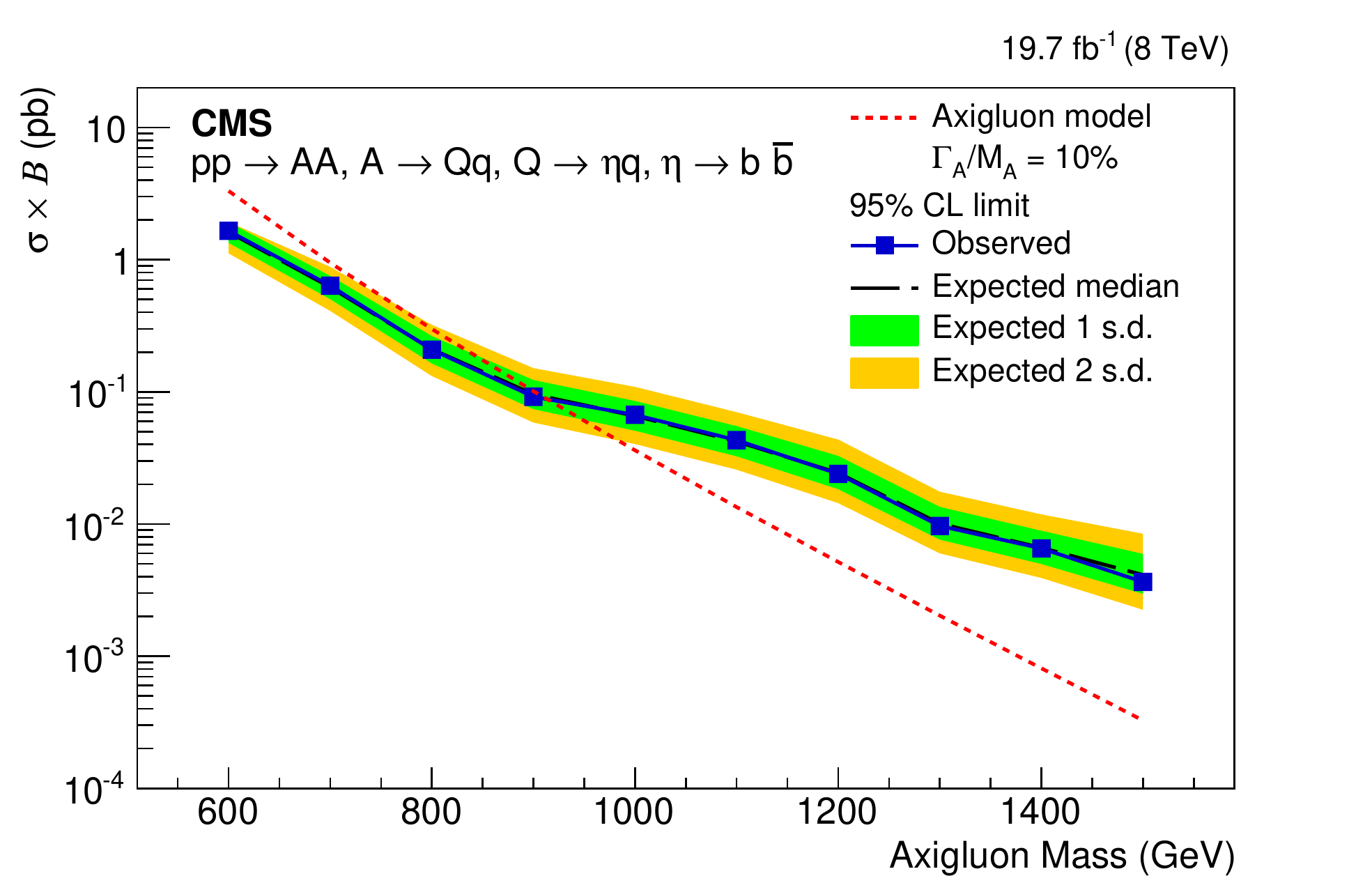}
  \caption{Upper limits at $95\%$ CL on signal cross section times branching fraction, as a function of axigluon mass $M_\mathrm{A}$, assuming a width of 3.5\% (\cmsLeft) and 10\% (\cmsRight) of $M_\mathrm{A}$, and a decay according to the A2 model. The observed cross section limits (points) are compared with the expected limit (dashed line) and the one and two standard deviation uncertainty bands. The cross section for axigluon pair production (dashed red line) is also shown.}
  \label{fig:LimitsAxiwithBs}
\end{figure}

Figure~\ref{fig:LimitsAxiwidth10-15} shows similar results for axigluon pair production and decay according to the A1 model. We exclude axigluon masses from 0.6 up to 1.15\TeV, depending on the model parameters. For axigluons decaying according to the A2 model, we exclude axigluon masses from 0.6 up to 0.9\TeV, as shown in Fig.~\ref{fig:LimitsAxiwithBs}.

\begin{figure*}[tbh]
  \centering
 \includegraphics[width=0.49\textwidth]{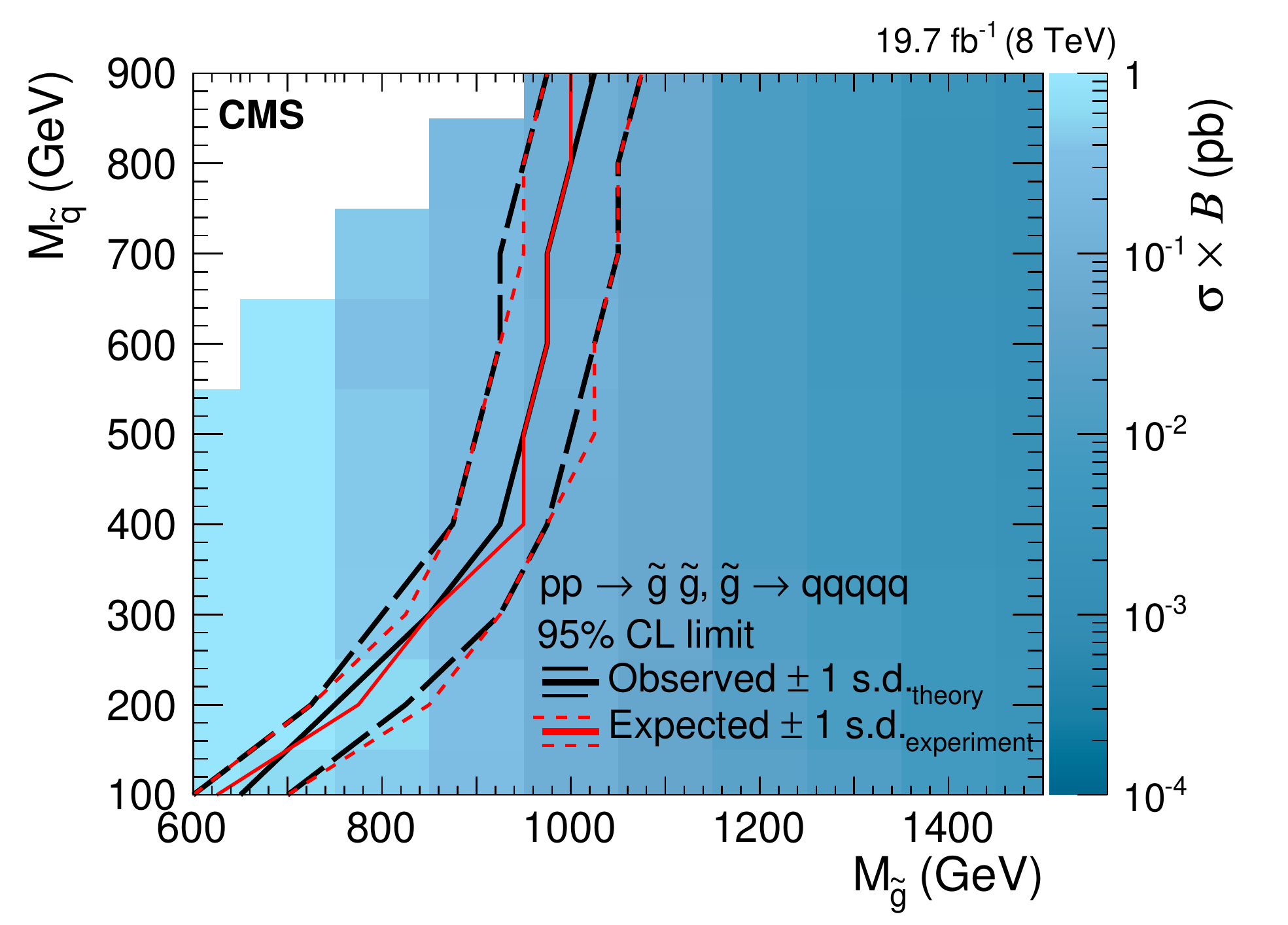}
  \includegraphics[width=0.49\textwidth]{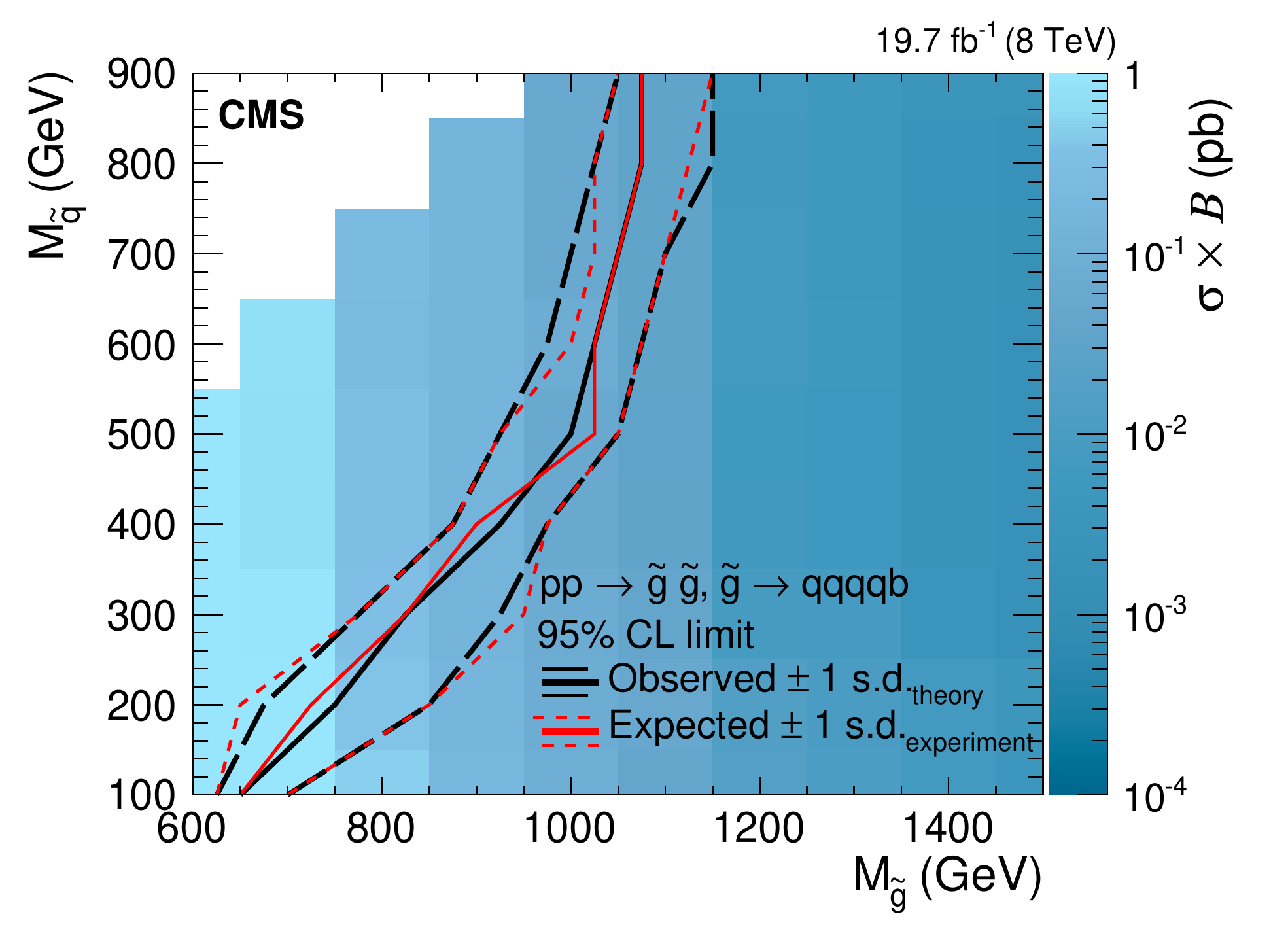}
  \includegraphics[width=0.49\textwidth]{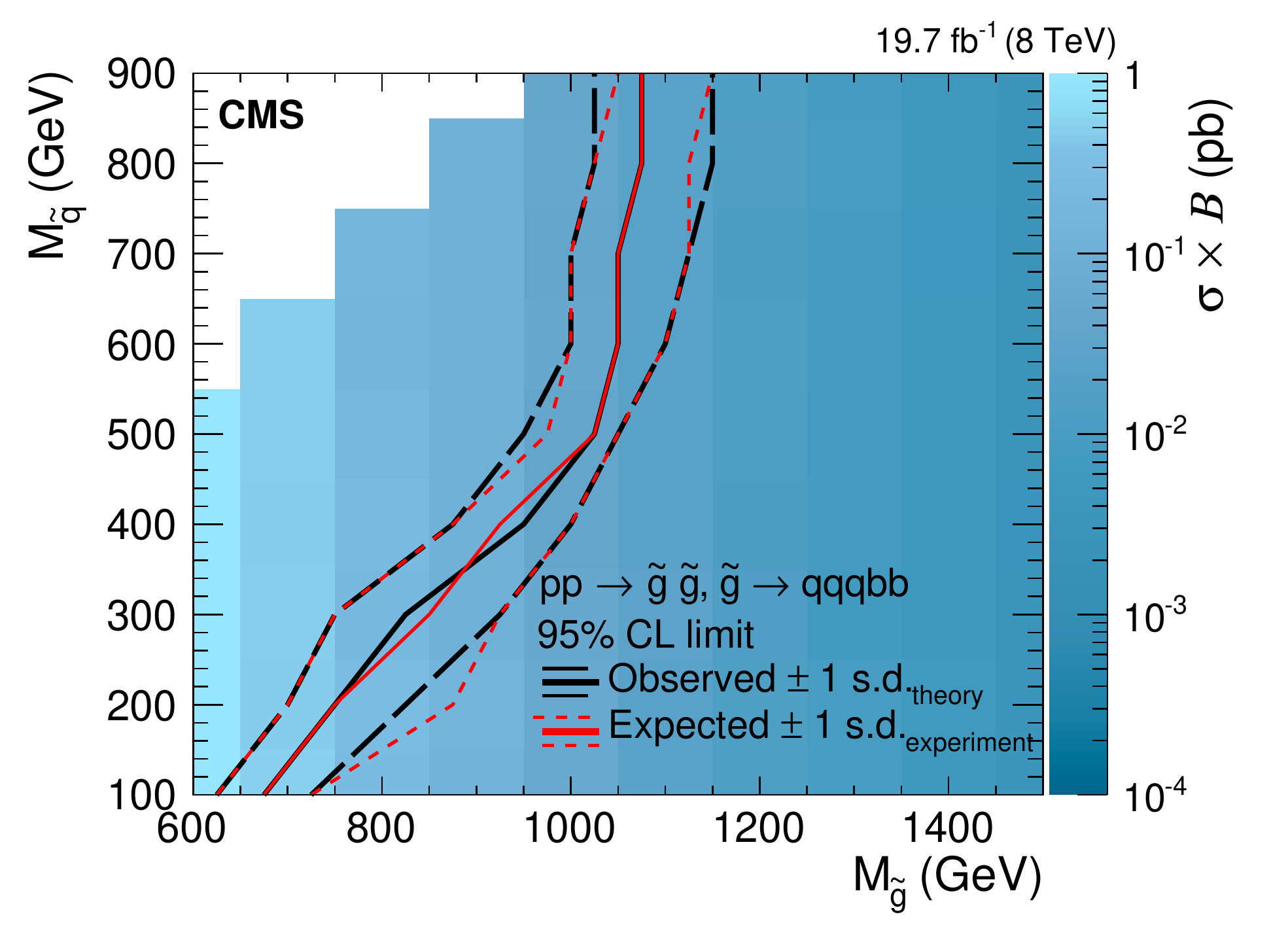}
  \includegraphics[width=0.49\textwidth]{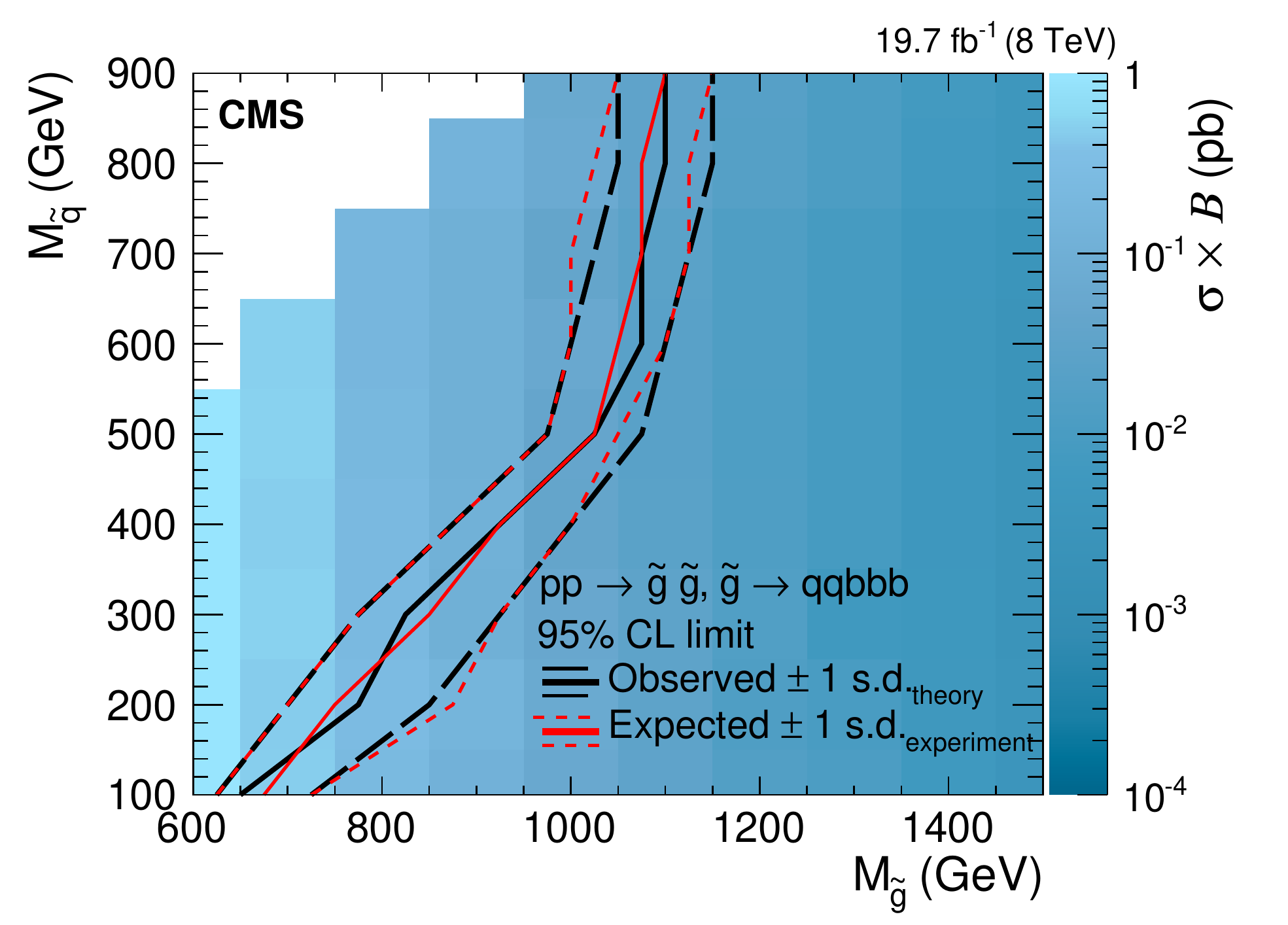}
 \caption{Upper limits at $95\%$ CL on the signal cross section times branching fraction, as a function of the gluino mass $M_{\sGlu}$ and squark mass $M_{\sQua}$ for the pair-produced gluino model with RPV decays in the final states qqqqq (G1, top left), qqqqb (G2, top right), qqqbb (G3, bottom left), and qqbbb (G4, bottom right). The observed limit (black lines) is compared to the expected limit (red lines) with the one standard deviation theoretical uncertainty in the observed limit (black dashed lines) and the one standard deviation statistical and systematic uncertainties combined in the expected limits (red dashed lines). The gluino pair production cross sections are shown with the color scale.}
  \label{fig:LimitsGluino10j}
\end{figure*}

We exclude gluinos with RPV decay G1 with masses from 0.6 up to 0.65--1.03\TeV for squark masses from 0.1 to 0.9\TeV, as shown in Fig.~\ref{fig:LimitsGluino10j} (top left). For the G2 decay mode, we exclude gluino masses from 0.6 up to 0.65--1.08\TeV for squark masses from 0.1 to 0.9\TeV, as shown in Fig.~\ref{fig:LimitsGluino10j} (top right). For the G3 decay mode we exclude gluino masses from 0.6 up to 0.68--1.08\TeV for the bottom squark masses from 0.1 to 0.9\TeV, as shown in Fig.~\ref{fig:LimitsGluino10j} (bottom left). Finally, for the G4 decay mode we exclude gluino masses from 0.6 up to 0.65--1.1\TeV for the bottom squark masses from 0.1 to 0.9\TeV, as shown in Fig.~\ref{fig:LimitsGluino10j} (bottom right).
\ifthenelse{\boolean{cms@external}}{}{\clearpage}
\section{Summary}

A search has been performed for pair production of color-octet vector boson resonances and of gluinos in an RPV SUSY model, in inclusive 8- and 10-jet final states. The search is based on data from proton-proton collisions at $\sqrt{s}=8\TeV$ corresponding to an integrated luminosity of 19.7\fbinv collected by the CMS experiment at the LHC. The scalar sum of the transverse momenta of the jets is used as a discriminating variable, with additional requirements placed on event sphericity and b-tagged jet multiplicity. The dominant QCD multijet background is estimated from control samples at lower multiplicity, without any reliance on simulation. No significant deviation from the standard model background predictions has been observed. Upper limits at $95\%$ confidence level on the cross section times branching fraction have been set for several signal scenarios. The cross section limits have been compared to specific coloron, axigluon, and gluino pair production cross sections. For the coloron and axigluon models, the lowest excluded mass is 0.6\TeV, while the highest excluded mass ranges from 0.75 to 1.15\TeV. For the RPV SUSY model, the lowest excluded mass is 0.6\TeV, while the highest excluded mass is 1.1\TeV. Models with colorons and axigluons decaying in multijet final states are probed experimentally for the first time.

\begin{acknowledgments}
We congratulate our colleagues in the CERN accelerator departments for the excellent performance of the LHC and thank the technical and administrative staffs at CERN and at other CMS institutes for their contributions to the success of the CMS effort. In addition, we gratefully acknowledge the computing centres and personnel of the Worldwide LHC Computing Grid for delivering so effectively the computing infrastructure essential to our analyses. Finally, we acknowledge the enduring support for the construction and operation of the LHC and the CMS detector provided by the following funding agencies: BMWFW and FWF (Austria); FNRS and FWO (Belgium); CNPq, CAPES, FAPERJ, and FAPESP (Brazil); MES (Bulgaria); CERN; CAS, MoST, and NSFC (China); COLCIENCIAS (Colombia); MSES and CSF (Croatia); RPF (Cyprus); SENESCYT (Ecuador); MoER, ERC IUT and ERDF (Estonia); Academy of Finland, MEC, and HIP (Finland); CEA and CNRS/IN2P3 (France); BMBF, DFG, and HGF (Germany); GSRT (Greece); OTKA and NIH (Hungary); DAE and DST (India); IPM (Iran); SFI (Ireland); INFN (Italy); MSIP and NRF (Republic of Korea); LAS (Lithuania); MOE and UM (Malaysia); BUAP, CINVESTAV, CONACYT, LNS, SEP, and UASLP-FAI (Mexico); MBIE (New Zealand); PAEC (Pakistan); MSHE and NSC (Poland); FCT (Portugal); JINR (Dubna); MON, RosAtom, RAS and RFBR (Russia); MESTD (Serbia); SEIDI and CPAN (Spain); Swiss Funding Agencies (Switzerland); MST (Taipei); ThEPCenter, IPST, STAR and NSTDA (Thailand); TUBITAK and TAEK (Turkey); NASU and SFFR (Ukraine); STFC (United Kingdom); DOE and NSF (USA).

Individuals have received support from the Marie-Curie programme and the European Research Council and EPLANET (European Union); the Leventis Foundation; the A. P. Sloan Foundation; the Alexander von Humboldt Foundation; the Belgian Federal Science Policy Office; the Fonds pour la Formation \`a la Recherche dans l'Industrie et dans l'Agriculture (FRIA-Belgium); the Agentschap voor Innovatie door Wetenschap en Technologie (IWT-Belgium); the Ministry of Education, Youth and Sports (MEYS) of the Czech Republic; the Council of Science and Industrial Research, India; the HOMING PLUS programme of the Foundation for Polish Science, cofinanced from European Union, Regional Development Fund, the Mobility Plus programme of the Ministry of Science and Higher Education, the National Science Center (Poland), contracts Harmonia 2014/14/M/ST2/00428, Opus 2013/11/B/ST2/04202, 2014/13/B/ST2/02543 and 2014/15/B/ST2/03998, Sonata-bis 2012/07/E/ST2/01406; the Thalis and Aristeia programmes cofinanced by EU-ESF and the Greek NSRF; the National Priorities Research Program by Qatar National Research Fund; the Programa Clar\'in-COFUND del Principado de Asturias; the Rachadapisek Sompot Fund for Postdoctoral Fellowship, Chulalongkorn University and the Chulalongkorn Academic into Its 2nd Century Project Advancement Project (Thailand); and the Welch Foundation, contract C-1845.
\end{acknowledgments}

\ifthenelse{\boolean{cms@external}}{\vspace*{4ex}}{}
\bibliography{auto_generated}

\cleardoublepage \appendix\section{The CMS Collaboration \label{app:collab}}\begin{sloppypar}\hyphenpenalty=5000\widowpenalty=500\clubpenalty=5000\input{EXO-13-001-authorlist.tex}\end{sloppypar}
\end{document}

%% file: EXO-13-001-authorlist.tex
\textbf{Yerevan Physics Institute,  Yerevan,  Armenia}\\*[0pt]
V.~Khachatryan, A.M.~Sirunyan, A.~Tumasyan
\vskip\cmsinstskip
\textbf{Institut f\"{u}r Hochenergiephysik der OeAW,  Wien,  Austria}\\*[0pt]
W.~Adam, E.~Asilar, T.~Bergauer, J.~Brandstetter, E.~Brondolin, M.~Dragicevic, J.~Er\"{o}, M.~Flechl, M.~Friedl, R.~Fr\"{u}hwirth\cmsAuthorMark{1}, V.M.~Ghete, C.~Hartl, N.~H\"{o}rmann, J.~Hrubec, M.~Jeitler\cmsAuthorMark{1}, A.~K\"{o}nig, I.~Kr\"{a}tschmer, D.~Liko, T.~Matsushita, I.~Mikulec, D.~Rabady, N.~Rad, B.~Rahbaran, H.~Rohringer, J.~Schieck\cmsAuthorMark{1}, J.~Strauss, W.~Treberer-Treberspurg, W.~Waltenberger, C.-E.~Wulz\cmsAuthorMark{1}
% ...
\vskip\cmsinstskip
% ...
1:~~Also at Vienna University of Technology, Vienna, Austria\\
% ...
